\documentclass[useAMS,usenatbib]{mn2e} 
\usepackage{aas_macros}
\usepackage{graphics}
\usepackage[pdftex]{graphicx}
\usepackage{epstopdf}
\usepackage{epsfig}  
\usepackage{natbib} 
\usepackage{float}
\usepackage{amsmath}
\usepackage{times}
\usepackage[varg]{txfonts}
\usepackage{verbatim} 
\usepackage{multirow,bigdelim} 
\usepackage{color}
\usepackage{vmargin}
\setmarginsrb{1.2cm}{1cm}{1.2cm}{1cm}{1cm}{1cm}{1cm}{1cm}

\newcommand{\hmsun}{{\,\rm h^{-1}M}_\odot}

  \makeatletter
    \renewcommand{\paragraph}{\@startsection{paragraph}{4}{\z@}%
      {-3.25ex\@plus -1ex \@minus -.2ex}%
      {1.5ex \@plus .2ex}%
      {\normalfont\small\centering}}
     
    \renewcommand{\subparagraph}{\@startsection{subparagraph}{5}{\z@}%
      {-3.25ex\@plus -1ex \@minus -.2ex}%
      {1.5ex \@plus .2ex}%
      {\normalfont\small\centering}}
    \makeatother

\newcommand{\ginnungagap}{{\sc Ginnungagap}}

\newcommand{\gadget}{{\sc Gadget}}

\newcommand{\kms}{{ km~s$^{-1}$}}
\newcommand{\hMpc}{{ $h^{-1}$~Mpc}}

\title[Constrained Simulations]{Cosmicflows Constrained Local UniversE Simulations}
\author[Sorce et al. ]
{Jenny G. Sorce$^{1}$\thanks{E-mail: \text{jsorce@aip.de}}, 
Stefan Gottl\"{o}ber$^1$,
Gustavo Yepes$^2$,
Yehuda Hoffman$^3$, 
Helene M. Courtois$^4$,
\and Matthias Steinmetz$^{1}$, 
R. Brent Tully$^5$,
Daniel Pomar\`ede$^6$, Edoardo Carlesi$^3$\\
$^1$Leibniz-Institut f\"{u}r Astrophysik, 14482 Potsdam, Germany\\
$^2$ Departamento de F\'{\i}sica Te\'orica, Universidad Aut\'onoma de Madrid, Cantoblanco, 28049, Madrid, Spain \\
$^3$Racah Institute of Physics, Hebrew University, Jerusalem, Israel\\
$^4$Universit\'e Claude Bernard Lyon I, Institut de Physique Nucl\'eaire, Lyon, France \\
$^5$ Institute for Astronomy, University of Hawaii, 2680 Woodlawn Drive, Honolulu, HI 96822, USA\\
$^6$Institut de Recherche sur les Lois Fondamentales de l'Univers, CEA Saclay, 91191 Gif-sur-Yvette, France\\
}

\begin{document}

\date{}

\pagerange{\pageref{firstpage}--\pageref{lastpage}} \pubyear{2015}

\maketitle

\label{firstpage}

\begin{abstract}
\indent 

This paper combines observational datasets and cosmological simulations to generate realistic numerical replicas of the nearby Universe. These latter are excellent laboratories for studies of the non-linear process of structure formation in our neighborhood. With measurements of radial peculiar velocities in the Local Universe ({\it cosmicflows-2}) and a newly developed technique, we produce Constrained Local UniversE Simulations (CLUES). To assess the quality of these constrained simulations, we compare them with random simulations as well as with local observations. The cosmic variance, defined as the mean one-sigma scatter of cell-to-cell comparison between two fields, is significantly smaller for the constrained simulations than for the random simulations. Within the inner part of the box where most of the constraints are, the scatter is smaller by a factor 2 to 3 on a 5 \hMpc\ scale with respect to that found for random simulations. This one-sigma scatter obtained when comparing the simulated and the observation-reconstructed velocity fields is only 104 $\pm$ 4 \kms\ i.e. the linear theory threshold. These two results demonstrate that these simulations are in agreement with each other and with the observations of our neighborhood. For the first time, simulations constrained with observational radial peculiar velocities resemble the Local Universe up to a distance of 150 \hMpc\ on a scale of a few tens of megaparsecs. When focusing on the inner part of the box, the resemblance with our cosmic neighborhood extends to a few megaparsecs ($<$ 5 \hMpc). The simulations provide a proper Large Scale environment for studies of the formation of nearby objects.
 
\end{abstract}


\section{Introduction}

The formation of structures in the Universe, from tiny fluctuations at the era of recombination to the large diversity observed today, is a highly non-linear process. Its multi-scale nature is best studied by numerical experimentation. Cosmological simulations of structure formation rely on the cosmological principle which assumes the homogeneity of the Universe on large enough scales. The random nature of the primordial Gaussian perturbation field however implies that separate patches of the Universe are not identical. In that context, properties of patches of a few megaparsecs on aside vary widely. To overcome this cosmic variance, statistical comparisons to study the formation of structures are based on large observational datasets \citep[e.g.][]{2002AJ....123..485S,2003AJ....126.2081A,2009ApJS..182..543A} and large cosmological simulations \citep[e.g.,][]{Klypin2011,DeusSimulation2012,Prada2012,AnguloXXL2012,2014MNRAS.437.3776W,2014arXiv1411.4001K,2014arXiv1407.2600S}. In order to also resolve small scale structures in the entire required large computational boxes, the mass resolution must be sufficiently high, resulting in simulations that can be time consuming and expensive. 

An alternative approach is to reduce the cosmic variance by focusing on the numerical study of structure formation in the nearby Universe and a direct comparison of theoretical models with local observations. There is a double advantage of producing simulations resembling the Local Universe. First, our neighborhood is without any doubt the best-observed volume of the Universe and as such hides its own cosmological treasures, often as fossils from its early epochs. Second, because a very large box size is not required for such simulations, the desired high resolution can be more easily achieved without being overly time consuming. However, this double advantage requires a correct modeling of the initial conditions based on the observed structures in the Local Universe. Standard cosmological simulations are obtained with initial conditions drawn from a random realization of the primordial perturbation field for a given cosmological model. Observational data of the Local Universe are used as additional constraints on these initial conditions so that resulting simulations resemble the Local Universe. The resulting constrained simulations attempt to describe the evolution of the observed structures in the nearby Universe.

The billions of data points that characterize the initial conditions of a cosmological simulation cannot be constrained by only thousands of observational data. The main aim of the different existing reconstruction techniques is to reduce the cosmic variance of the resulting constrained simulations. Local object candidates can then be identified to study their formation and evolution in the proper environment. Since the introduction of the POTENT method and the first attempt to reconstruct initial conditions from sparse observational data of the velocity field traced by galaxies \citep{1990ApJ...364..349D,1990ApJ...364..370B,1992ApJ...391..443N} a lot of progress has been made over the last two and half decades. The constrained realization technique proposed by \citet{1991ApJ...380L...5H} constitutes such a significant step forward in developing further the reconstruction method. With this technique, \citet{1993ApJ...415L...5G} have constructed constrained initial conditions using the POTENT data that led to the first $N$-body simulation that mimicked the matter distribution around the Local Group in a 256 \hMpc\ box \citep{1996ApJ...458..419K}.

The first step in generating initial conditions for constrained simulations consists in reconstructing today's three-dimensional density field from sparse and noisy observational data of the local galaxies. These observational data can be either the positions and radial peculiar velocities of galaxies \citep[]{ 2002ApJ...571..563K,2003ApJ...596...19K, 2014MNRAS.437.3586S} or redshift catalogs \citep[]{2013MNRAS.435.2065H}. In a second step the initial linear density field must be retrieved to derive the initial conditions for the simulations. There are two possible ways: backwards as described is section 2 of this paper or forwards as recently proposed by \citet{2013MNRAS.429L..84K,2013MNRAS.435.2065H,2013MNRAS.432..894J, 2013ApJ...772...63W}. In the latter case, the initial density field is sampled from a probability distribution function consisting of a Gaussian prior and a likelihood. A complete overview about the different methods is given in \cite{2014ApJ...794...94W}. 

Coming back to the observational dataset used here to constrained the initial conditions, this paper is part of the CLUES project\footnote{http://www.clues-project.org/} \citep[Constrained Local UniversE Simulations,][]{2010arXiv1005.2687G, 2014NewAR..58....1Y} which focuses on using peculiar velocities as constraints. Within this project, indeed, a number of constrained simulations of the Local Universe, using peculiar velocities from \citet{2004AJ....127.2031K,1997ApJS..109..333W,2001ApJ...546..681T} as observational constraints, have been run. Although estimating peculiar velocities constitutes an observational challenge, there is a double advantage in using them: first they are highly linear and correlated on large scales ; second, they are excellent tracers of the underlying gravitational field as they account for both the baryonic and the dark matter. 

However, the first generation of these constrained simulations was affected by a substantial shift in the positions of objects recovered at redshift zero. To reduce this shift a new technique has been developed to account for the cosmic displacement field \citep{2013MNRAS.430..902D,2013MNRAS.430..912D,2013MNRAS.430..888D} and the noisy velocity information \citep{2014MNRAS.437.3586S}. The method has been applied to the first catalog of the Cosmicflows project\footnote{http://www.ipnl.in2p3.fr/projet/cosmicflows/} \citep{2008ApJ...676..184T} producing simulations resembling the Local Universe down to a few megaparsecs within 30 \hMpc\ \citep{2014MNRAS.437.3586S}.  

At the end of 2013 \citep{2013AJ....146...86T}, a second catalog of the Cosmicflows project was released. Superior in size (number of constraints, $\sim$ 8000 against 2000) and extent ($\sim$ 150 \hMpc\ against 30 \hMpc) to the first catalog, it constitutes an ideal supplier of observational data to construct initial conditions constrained by the observed peculiar velocities of galaxies. Uniting the second release of the Cosmicflows data and the newly developed technique to build more accurate constrained initial conditions, this paper aims at demonstrating by how much the cosmic variance is decreased with respect to random simulations and how far the resemblance with the Local Universe can be extended. 

The structure of the paper is as follow. In the second section, we briefly review the methodology described in \citet{2014MNRAS.437.3586S} and \citet{2015MNRAS.450.2644S} which combines several techniques to reduce biases in the observational catalog and to build constrained initial conditions. The third section presents the resulting constrained simulations of the Local Universe: the cosmic variance is estimated,  the Large Scale Structure is compared with galaxies from the 2MASS redshift catalog \citep{2012ApJS..199...26H} and with the three-dimensional reconstruction of the overdensity and velocity fields of the Local Universe at redshift zero. In addition, the simulated Laniakea Supercluster of galaxies, a basin of attraction of local velocity flows, is compared with the observed Laniakea Supercluster \citep{2014Natur.513...71T}. Finally, before concluding, dark matter halo candidates for well known nearby clusters, such as Virgo, are identified in the simulations.


\section{Methodology}

In this section we discuss how the initial conditions have been constructed from the observational data: a set of positions and radial velocities of galaxies. Important steps in this procedure are the grouping of galaxies and the minimization of biases. 

\subsection{Observational data: {\it Cosmicflows-2}}

{\it Cosmicflows-2} is the second generation observational catalog of galaxy distances built by the Cosmicflows collaboration. Published by \citet{2013AJ....146...86T}, it contains more than 8,000 accurate galaxy distances. Distance measurements come mostly from the Tully-Fisher relation \citep{1977A&A....54..661T} and the Fundamental Plane methods \citep{2001MNRAS.321..277C}. Cepheids \citep{2001ApJ...553...47F}, Tip of the Red Giant Branch \citep{1993ApJ...417..553L}, Surface Brightness Fluctuation \citep{2001ApJ...546..681T}, supernovae of type Ia \citep{2007ApJ...659..122J} and other miscellaneous methods also contribute to this large dataset but to a minor extent ($\sim$ 12\%) although they have individually higher weights because of smaller errors.

The final goal of the paper is to build initial conditions constrained by this catalog in order to run cosmological simulations, working above the scale of galaxy virial motions in clusters (non-linear displacements) is required. Therefore, in this paper, the grouped version of {\it cosmicflows-2}  is used. With a method similar to that described in \citet{2015AJ....149...54T,2015AJ....149..171T}, 552 groups and 4303 single galaxies can be identified in the dataset shrinking the number of constraints to 4855. The resulting number density is large enough to construct constrained initial conditions as demonstrated by \citet{2013MNRAS.430..902D,2013MNRAS.430..912D,2013MNRAS.430..888D}. 

\begin{figure}
\hspace{-0.5cm}\includegraphics[width=0.5\textwidth]{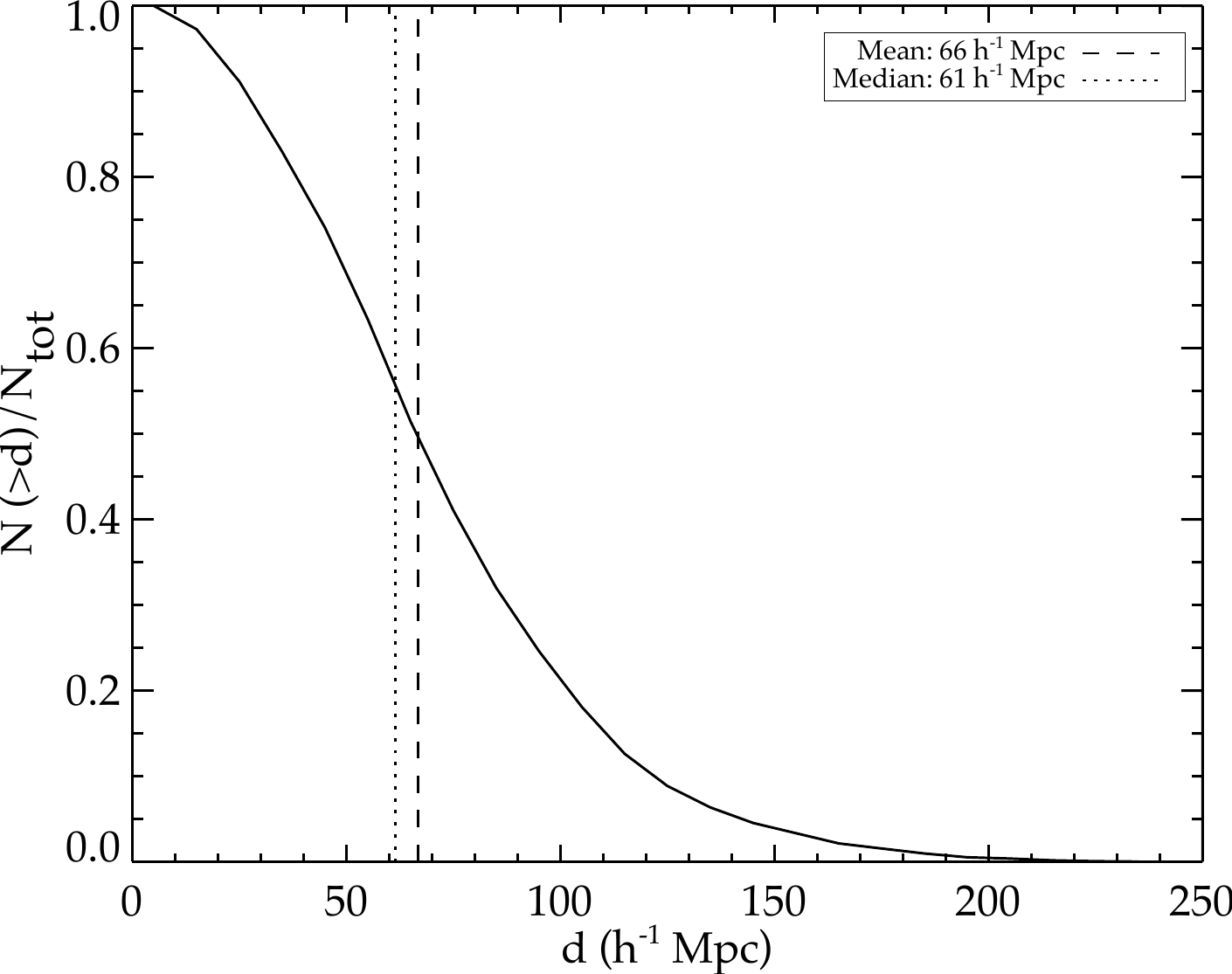}
\caption{Normalized cumulative distribution of datapoint distances in {\it cosmicflows-2} (solid line). The mean and median distances, at 66 and 61 \hMpc\ respectively, are marked by dashed and dotted black lines.}
\label{fig:cumuldistri}
\end{figure}

Figure \ref{fig:cumuldistri} shows the normalized cumulative distribution of distances in {\it cosmicflows-2} as well as the mean and median distances. The catalog extends up to 230 \hMpc\ but on average constraints are within $\sim$ 66 \hMpc\  ($\sim$ 61 \hMpc\ for the median). In fact, about 60 \% of the constraints are within a sphere of 70 \hMpc\ radius, approximately 80 \% are within 100 \hMpc, and $\sim$ 98 \% are within 160 \hMpc. From this distribution, the inner box is expected to be the most constrained part of the simulation and beyond 160 \hMpc\ the random component is anticipated to overcome the constraints. Intermediate results should be found in between.

\subsection{Minimizing Biases in the Observational Catalog}

\citet{2013AJ....146...86T} warned us that {\it cosmicflows-2} is affected by biases with effects that cannot be ignored anymore as the effects are stronger with increasing distance. 
There are four biases known: 
\begin{itemize}
\item In the literature the first bias has been given a number of terms that are used interchangeably. These are Problem I, Selection Effect/Bias, r against V, Distance-dependent, Frequentist, Calibration problem, M-bias of the second kind  \citep[Kaptney, 1914; Malmquist, 1922;][]{1992ApJ...395...75H,1994ApJ...430....1S,1997ARA&A..35..101T,1993A&A...280..443T,1990A&A...234....1T,1994ApJ...435..515H,1994ApJS...92....1W,1995ApL&C..31..263T}. This bias is analogous to a selection effect in magnitude (dim galaxies are selectively excluded from the observational sample) resulting in underestimated distances. This bias has been nulled in the observational {\it cosmicflows-2} catalog \citep[e.g.][]{2012ApJ...749...78T,2013ApJ...765...94S,2014MNRAS.444..527S}. 
\item The second bias, referred to as Homogeneous Malmquist Bias or Problem II, General Malmquist Bias, Geometry Bias, V against r, Classical, Bayesian, Inferred-distance problem, M-bias of the first kind in the literature \citep[Kaptney, 1914; Malmquist, 1920;][]{1988ApJ...326...19L,1992ApJ...395...75H,1997ARA&A..35..101T,1994ApJ...430....1S,1993A&A...280..443T,1990A&A...234....1T,1994ApJ...435..515H,1995ApL&C..31..263T,1995PhR...261..271S}, is due to the fact that the number of observable galaxies from our perspective increases with the distance. There are more galaxies available to scatter inward due to errors than outward, creating the tendency to locate galaxies closer than they should be, namely to underestimate distances. 
\item In addition, because the Universe is inhomogeneous on small scales, galaxies are more likely to be scattered from high density regions towards low density regions than the opposite resulting in the Inhomogeneous Malmquist Bias \citep[e.g.][]{1994ARA&A..32..371D,1994MNRAS.266..468H,1992ApJ...391..494L}. 
\item On top of this last bias, there is a lognormal error distribution or an asymmetry in the distribution of fractional errors on distances because distances are derived from distance moduli via a logarithmic function. Thus if a same galaxy is located farther than it should be rather than closer, the error on the distance is larger although this is not reflected by the assigned fractional errors.
\end{itemize}

These biases lead mainly to a major infall onto the Local Volume. An iterative method to minimize the infall and reduce spurious non-gaussianities in the radial peculiar velocity distribution was applied to obtain a new distribution of radial peculiar velocities and corresponding distances
\citep{2015MNRAS.450.2644S}.

\subsection{Building Initial Conditions}

To build initial conditions for dark matter only numerical simulations (a set of particles with velocities and positions at a starting redshift) constrained by peculiar velocities, we rely on four techniques assuming a prior cosmological model:

\begin{itemize}
\item The Wiener-Filter method \citep[WF,][]{1995ApJ...449..446Z} to reconstruct the cosmic displacement field required to account for the displacement of constraints from their precursors' locations, 
\item the Reverse Zel'dovich Approximation (RZA) to relocate constraints at the positions of their progenitors \citep{2013MNRAS.430..888D,2013MNRAS.430..902D,2013MNRAS.430..912D} and to replace noisy radial peculiar velocities by their 3D reconstructions \citep{2014MNRAS.437.3586S},
\item the Constrained Realization \citep[CR,][]{1991ApJ...380L...5H} of Gaussian field technique to produce overdensity fields constrained by observational data, adding a random realization to compensate for the missing power spectrum. These latter are then converted into white noise that can be used to increase the resolution, 
\item the resolution is increased by adding some random small scale features in the white noise, then the white noise is converted back to build initial conditions for cosmological simulations.
\end{itemize}

These four steps can be summarized in a set of equations:
\begin{equation}
v^{WF}_\alpha(\mathbf{r})=\sum_{i=1}^n \langle v_\alpha(\mathbf{r})C_i\rangle \sum_{j=1}^n \langle C_i C_j\rangle^{-1}C_j \quad \mathrm{and} \quad \boldsymbol\psi^{\ WF}=\frac{\mathbf{v}^{\ WF}}{H_0\ f(t_{init})}
\label{eq:WF}
\end{equation}
where $\alpha$ = x, y, z and C$_i$ are the constraints plus their uncertainties and $f(t)=\frac{d\ (ln\ D(t))}{d\ (ln\ a(t))}$ is the growth rate (D the growth factor and a the scale factor). Brackets denote correlation functions depending solely on the assumed prior cosmological model (here the power spectrum) ; $\mathbf{v}$ is the velocity field and $\boldsymbol\psi$ is the displacement field. The 'WF' exponent means that a field is obtained with the Wiener-Filter method, 
\begin{equation}
\mathbf{x}_{init}^{\ RZA}=\mathbf{r} -\boldsymbol\psi^{\ WF} \quad \mathrm{and} \quad \mathbf{v}_{pec}=\mathbf{v}^{\ WF}
\label{eq:rza}
\end{equation}
where $\mathbf{x}_{init}^{\ RZA}$ is the approximate location of the constraints' precusors (linear theory at first order valid down to 2 \hMpc), $\mathbf{r}$ is the measured position of the constraints. Note that the left of Eq. \ref{eq:rza} is an extension of equation 14 in \citet{1991ApJ...379....6N}.
 \begin{equation}
 \begin{split}
\delta^{CR}(\mathbf{r})=\delta^{RR}(\mathbf{r})+\sum_{i=1}^n \langle\delta(\mathbf{r})C_i\rangle \sum_{j=1}^n \langle C_i C_j\rangle^{-1}(C_j-\overline{C_j})\\
\mathbf{\nabla . v}(t_{init}) =-\dot a f\delta^{CR} \qquad \qquad \qquad  \qquad \qquad \qquad \quad\;\;\,
\end{split}
\label{eq:CR}
\end{equation}
where $\delta$ stands for overdensity fields. $\delta^{RR}$ stands for the random realization field and each $\overline C_j$ represents a random constraint drawn from $\delta_{RR}$. $\delta^{CR}$ represents the constrained realization field.
\begin{equation}
\delta^{CR}(\mathbf{k})=\sqrt{P(\mathbf{k})\, V}\omega(\mathbf{k})
\end{equation}
where $\omega(k)$ stands for the white noise in Fourier space, $P$ represents the power spectrum, n is the number of particles and V the volume of the simulated box. This last equation is first used in reverse to convert the overdensity fields in white noises, then random small scale features are added to increase the resolution and the equation is used again to obtain the higher resolution density fields to prepare the initial conditions.

\begin{figure}
\vspace{-1cm}
\hspace{-0.6cm}\includegraphics[width=0.55\textwidth]{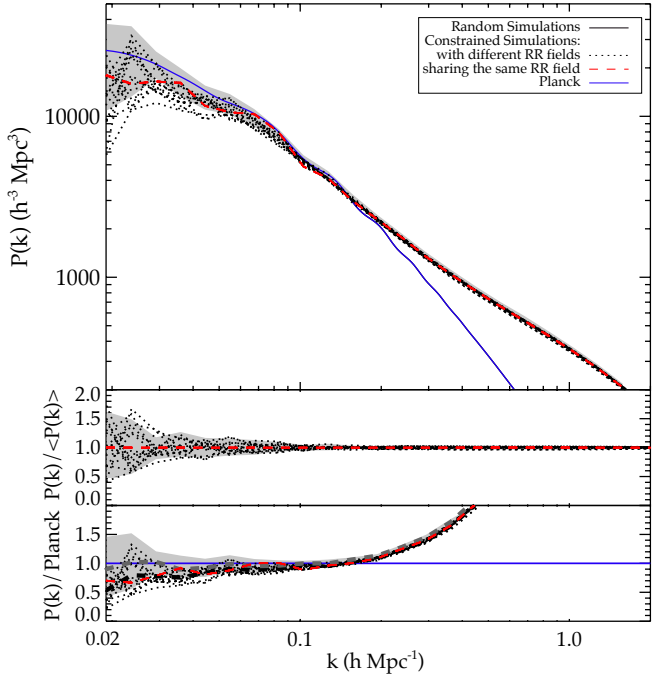}
\caption{Top: Power Spectra of 25 constrained simulations (15 constrained simulations with 15 different random realization fields in dotted black and 10 constrained simulations sharing the same random realization field but differing on the added small scale features in dashed red)  and 15 random simulations (grey area). The linear Planck power spectrum, used to build initial conditions, is represented by a solid blue line. Middle: Power Spectra divided by their respective mean. Bottom: Power Spectra divided by the Planck power spectrum. The dashed thick black and grey lines are the respective mean values. In the middle and bottom panels, the colors are the same as in the top panel.}
\label{fig:powspec}
\end{figure}

We apply the whole scheme to the observational catalog {\it cosmicflows-2} within the framework of Planck cosmology \citep[$\Omega_m$=0.307, $\Omega_\Lambda$=0.693, H$_0$=67.77, $\sigma_8=0.829$,][]{2014A&A...571A..16P}.


\section{Constrained Local UniversE Simulations}

In the second section, we noted that the largest distance of the {\it cosmicflows-2} catalog is 230 \hMpc, thus the size of the computational box should be sufficiently large to avoid spurious effects due to periodic boundary conditions where the observational data still have a constraining power. Tests have shown that a 500 \hMpc\ box meets this requirement. Such a computational volume extends the study up to the Shapley supercluster (the location of the farthest constraint) and with $512^3$  particles, the mass resolution is $8 \times 10^{10} \hmsun$ sufficient to resolve large groups and clusters of galaxies. Two of the simulations were also run at higher resolutions (1024$^3$ particles) to check that none of the conclusion drawn in this paper are affected by the 512$^3$ choice. As none of them are, we settle with the reasonable choice of 512$^3$ particles. A total of 25 constrained simulations and 15 random simulations have been performed in order to study the residual cosmic variance. 

\begin{figure*}
\vspace{-.5cm}
\includegraphics[width=0.46\textwidth]{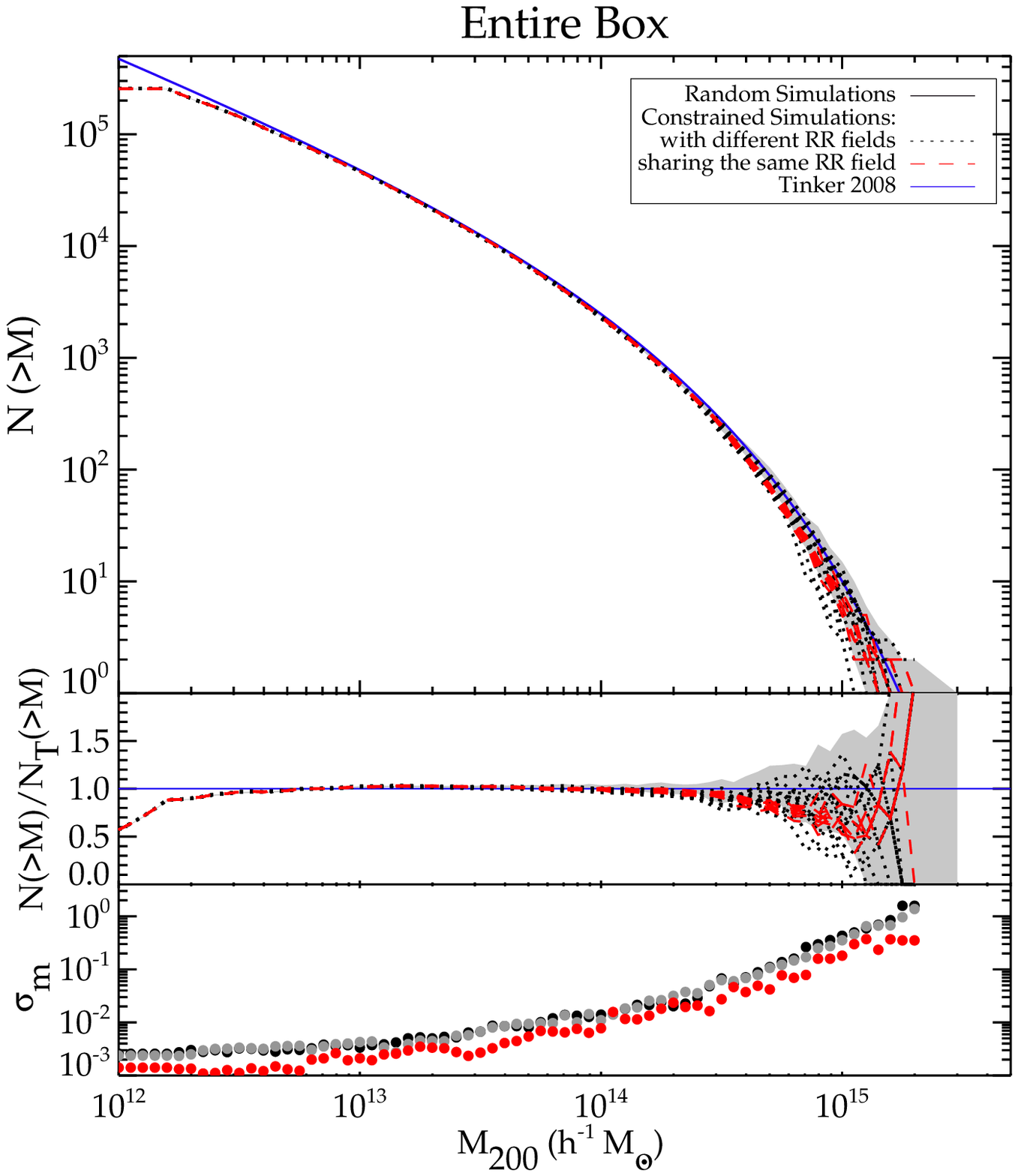}
\includegraphics[width=0.46\textwidth]{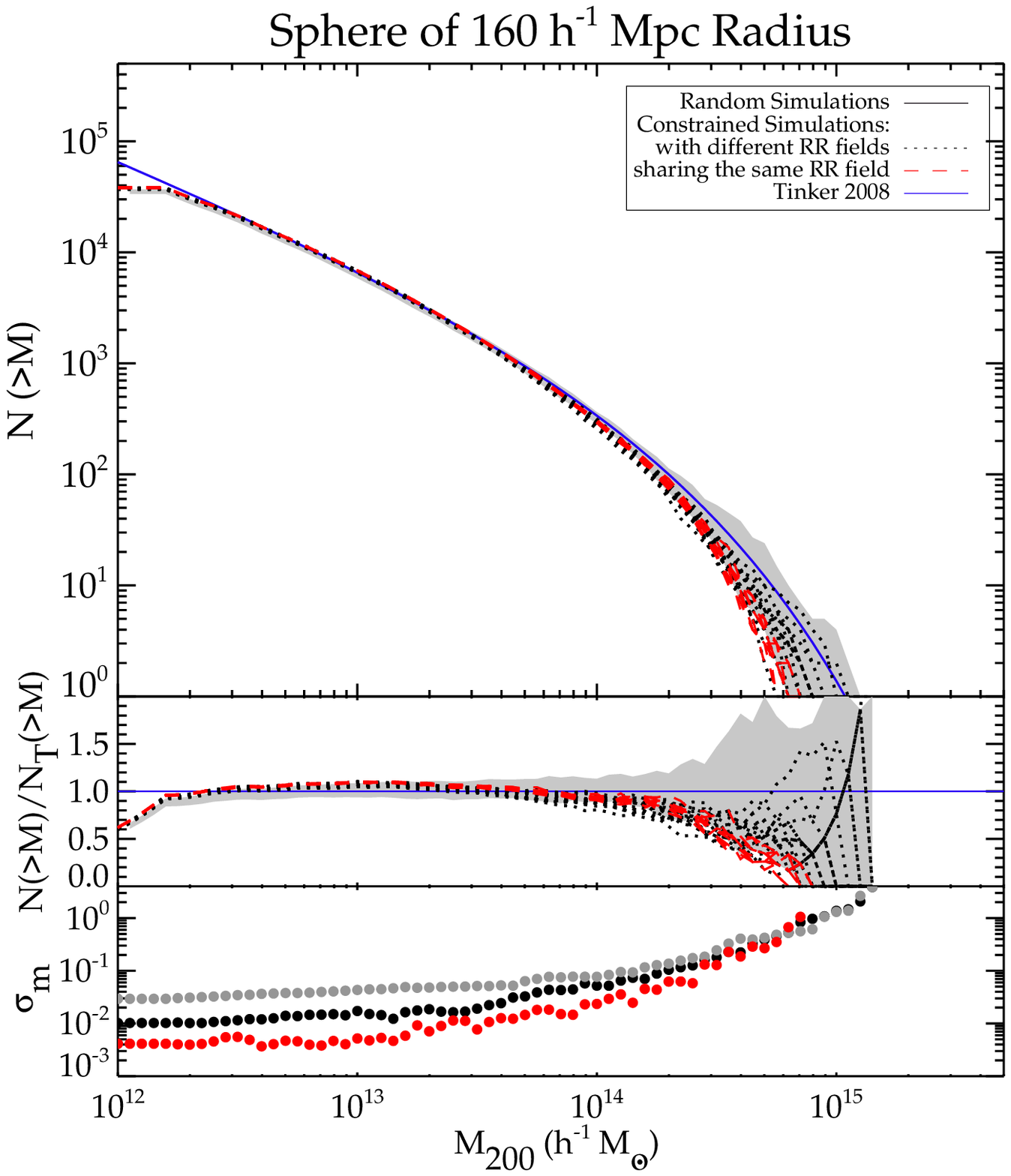}
\caption{Top: Cumulative Mass Functions of 25 constrained simulations (15 constrained simulations with 15 different random realization fields in dotted black and 10 constrained simulations sharing the same random realization field but differing on the added small scale features in dashed red), 15 random simulations (grey area) and Tinker cumulative mass function for Planck cosmology (solid blue line). Cumulative mass functions are derived with M$_{200}$ defined with respect to the critical density. Middle: Cumulative mass functions divided by the Tinker cumulative mass function, same color code as in the top plot. Bottom: Cumulative mass function scatters around their respective mean. Left: Cumulative mass functions for the entire 500 \hMpc\ box. Right: Cumulative mass functions for a 160 \hMpc\ radius sphere centered on the original box (where the Milky-Way-like is assumed to be). }
\label{fig:mass}
\end{figure*}

The initial conditions of these different simulations have been constructed in various ways:
\begin{itemize}
\item The first fifteen constrained initial conditions are built out of different random realization fields $\delta^{RR}$ plus the observational dataset (see equations \ref{eq:WF} to \ref{eq:CR}). This first step uses the Wiener Filter, the reverse Zel'dovich approximation and the constrained realization technique \citep{1999ApJ...520..413Z,2013MNRAS.430..888D,2014MNRAS.437.3586S,1992ApJ...384..448H} to generate $256^3$ density grids in accordance with the expected minimum scale on which the constraints are effective. The corresponding white noise field is used as an input to increase the resolution to $512^3$ particles with the \ginnungagap\ code\footnote{https://github.com/ginnungagapgroup/ginnungagap
}. This full `MPI+OpenMP' parallel code adds random small scale fluctuations in real space to increase the resolution to any level within a given cosmology. The resolution limit is dictated only by the total memory of the supercomputer. A final simulation is then characterized by two random seeds: the random realization field $\delta^{RR}$ and the added small scale features ;   
\item The additional ten constrained initial conditions share the same random realization field $\delta^{RR}$, but different seeds are used to increase the resolution to 512$^3$. They are thus expected to differ only on scales smaller than that of the input white noise field ;
\item Finally a set of fifteen random, i.e. not constrained, initial conditions has been constructed. They share the same seeds (same $\delta^{RR}$) as the fifteen constrained initial conditions. 
\end{itemize}

The simulations based on all these initial conditions have been performed with \gadget-3 \citep{2005MNRAS.364.1105S} from redshift 60 to redshift 0 with a 25 $h^{-1}$~kpc force resolution.\\

In Figure \ref{fig:powspec}, we first compare the resulting power spectra at $z=0$ to the linear power spectrum of the chosen cosmology. As expected the 15 random simulations (grey area) scatter at large scales around the linear input power spectrum (blue solid line). The 15 constrained simulations (dotted black line) tend to have less power on large scales, an effect which decreases with the box size as shown in the Appendix, because a smaller and smaller fraction of the box is constrained. The bottom panel of Figure 2 represents the power spectra divided by the Planck power spectrum as well as the mean values. Although on the low side, the power spectra of constrained simulations are within the scatter obtained with those of random simulations. Their mean is on average smaller by a factor 1.3 (factor that decreases with the scale) than that of random simulations on the large scales. As for the ten constrained simulations built out of the same random realization field, they share the same Large Scale power spectrum (red dashed lines) in agreement with the fact that the random features added to increase the resolution affect only the small scales. 
In the middle panel, the ratio of the power spectra to their mean for the three samples (15 constrained simulations, 10 constrained simulations with the same seed in $256^3$ and 15 random simulations) is displayed. By construction this is indistiguishable from a straight line for the 10 simulations where only small scale structures have been added. 

Next, the Amiga's Halo Finder \citep{2009ApJS..182..608K} is applied to each simulation to compile a list of dark matter halos. In Figure \ref{fig:mass}, using M$_{200}$ defined with respect to the critical density, the cumulative mass functions of the different simulations are compared with the same color code as that of Figure \ref{fig:powspec}. The blue color now stands for the Tinker cumulative mass function as defined by \citet{2008ApJ...688..709T} using the online mass calculator of \citet{2013A&C.....3...23M} and the Planck cosmological parameters as defined in section 2 of this paper. In the left panel of the figure, the cumulative mass functions for the entire 500 \hMpc\ boxes are shown (top), as well as their cumulative mass functions divided by the Tinker cumulative mass function (middle) and their scatters around their respective mean (bottom). The cumulative mass functions of the different simulations overlap and, as expected a smaller scatter is observed for the constrained simulations sharing the same random realization field (in red). In order to evaluate the effect of constraints on the cumulative mass functions, these latter are derived in a sphere of radius 160 \hMpc\ centered on the original box, namely on the center of the box. A radius of 160 \hMpc\ is a reasonable choice as it encompasses $\sim$ 98\% of the constraints (see section 2). 
The corresponding cumulative mass functions are shown in the right panel of Figure \ref{fig:mass}.  One can clearly see that at high mass ranges the cumulative mass functions of the constrained simulations are on the low side compared to the cumulative mass functions of the random realizations computed in the same sphere. These latter, contrary to the former, tend to scatter symmetrically around the Tinker cumulative mass function. As expected the scatter of the cumulative mass functions in the sphere is smaller for constrained simulations compared to random simulations (bottom of the right panel of Figure \ref{fig:mass}) and is the smallest for constrained simulations sharing the same random realization field. In this respect the cosmic variance is reduced in constrained simulations with respect to random simulations. We investigate more thoroughly the residual of the cosmic variance in the following subsection.

\subsection{Reduced Cosmic Variance}

\begin{figure}
\centering
\hspace{-.8cm}\includegraphics[width=0.5 \textwidth]{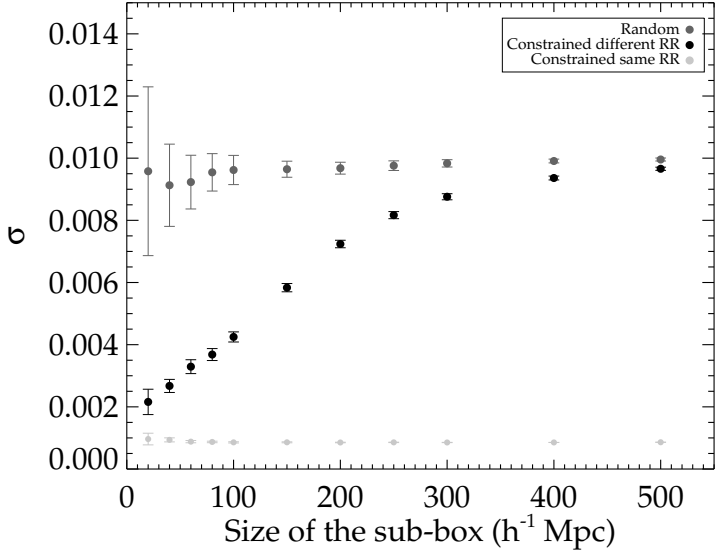}\\
\caption{Mean (circles) and scatter (error bars) of one-sigma scatters (standard deviations) obtained when deriving cell-to-cell comparisons of the initial (at starting redshift) density fields normalized by the mean density and smoothed on a 5 \hMpc\ scale as a function of the sub-box size. Cell-to-cell comparisons are conducted on pairs of simulations of the same set: random, dark grey - constrained, black - constrained sharing the same random realization field, light grey.}
\label{fig:c2c}
\end{figure}

\begin{figure}
\centering
\hspace{-1.cm}\includegraphics[width=0.5 \textwidth]{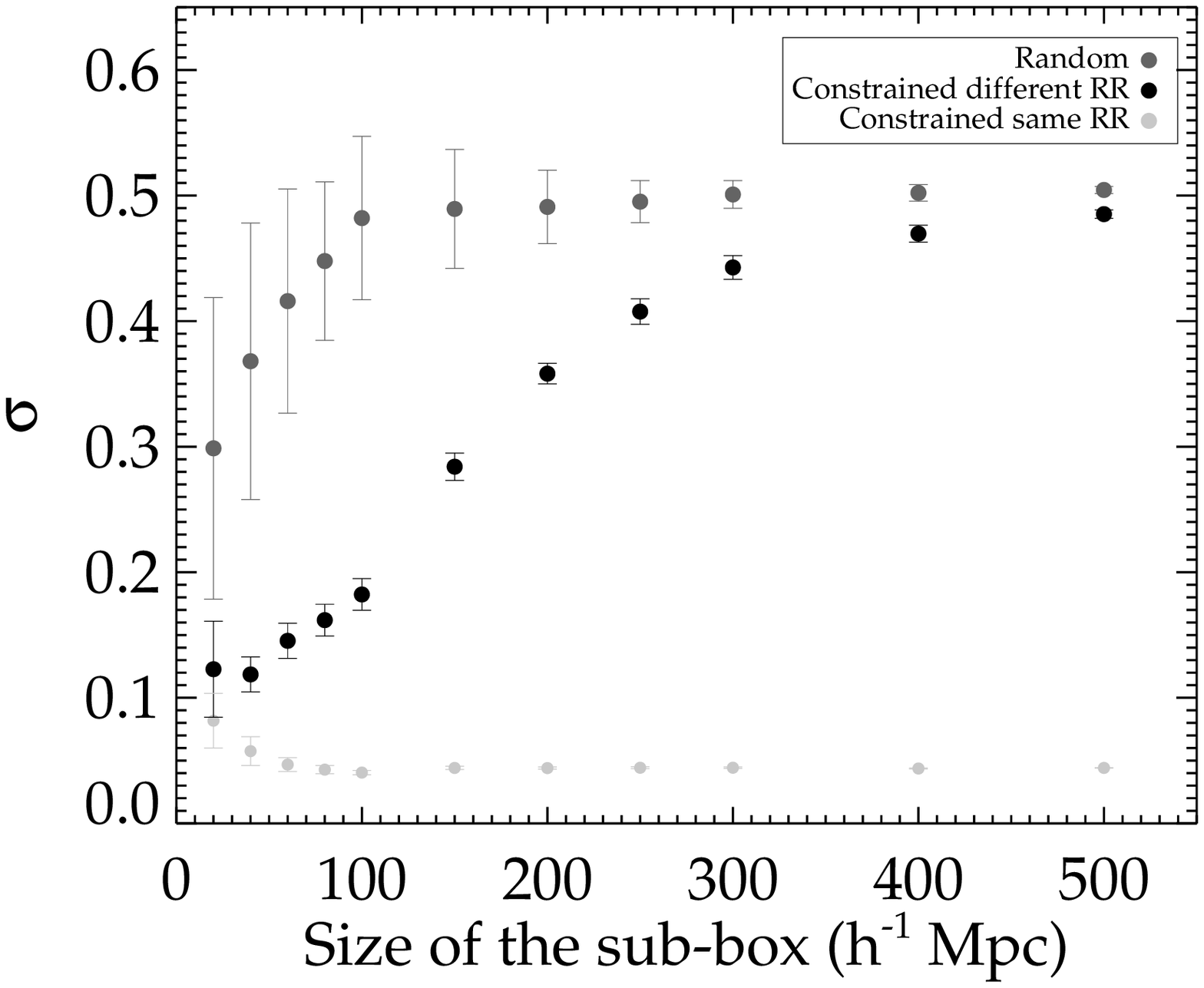}\\
\hspace{-1.3cm}\includegraphics[width=0.5 \textwidth]{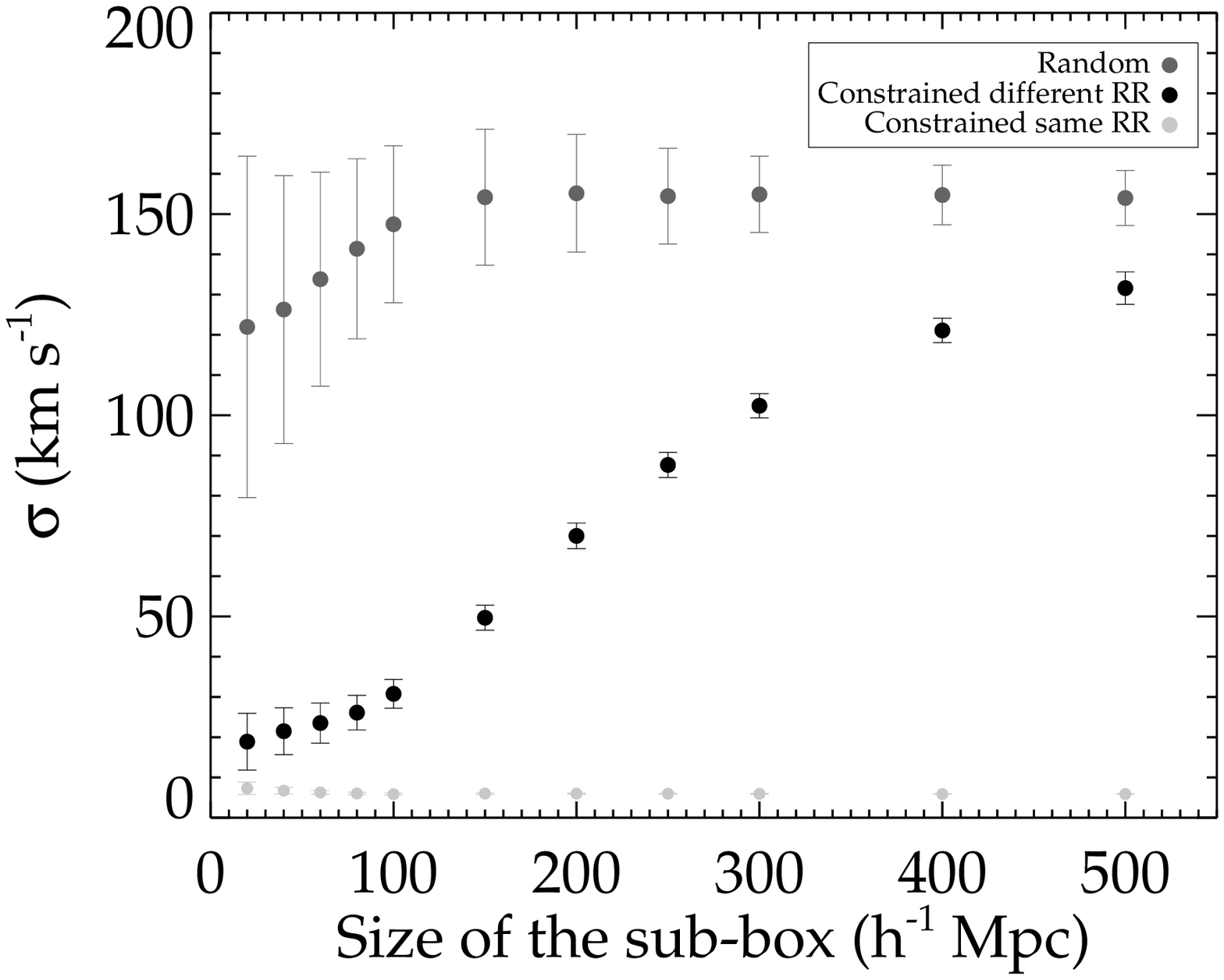}\\
\caption{Mean (circles) and scatter (error bars) of one-sigma scatters (standard deviations) obtained with cell-to-cell comparisons carried out on pairs of density fields normalized by the mean density (top) and of velocity fields (bottom) smoothed on a 5 \hMpc\ scale at z=0. Scatters are given as a function of the sub-box size. Pairs are constituted of two random (dark grey), two constrained (black) or two constrained, sharing the same random realization field, (light grey) simulations.}
\label{fig:c2c2}
\end{figure}

In this section we compare and quantify the cosmic variance within the different sets of simulations, i.e. constrained and random. To this end, a cloud-in-cell scheme on a $512^3$ grid is applied to the particle distributions of the initial conditions (z=60) and of the simulations at z=0 with a subsequent Gaussian smoothing on a scale of 5 \hMpc. Normalized by the mean density, the resulting smoothed density fields of any pair of constrained and random simulations are compared cell by cell. For each pair we build a density-density plot (the density field of a first simulation versus the density field of a second simulation). If the two simulations were identical all points would follow the 1:1 relation. We define the cosmic variance between two simulations as the one-sigma, hereafter 1$\sigma$, scatter (or standard deviation) around this 1:1 relation. We repeat this procedure for the 105 pairs of the 15 random simulations and those of the 15 constrained simulations as well as the 45 pairs of the 10 constrained simulations. Then we calculate the mean and the variance of the 1$\sigma$ scatters. The result is three points (filled dark grey, black and light grey circles for each one of the simulation types: random, constrained, constrained sharing the same random realization) in each of Figures \ref{fig:c2c} and \ref{fig:c2c2}, with error bars at the x-axis value of 500 \hMpc\ at z=60 and z=0 respectively. We repeat this procedure in smaller sub-boxes of size 400, 300, 250 \hMpc, etc, centered on the original box in order to measure the cosmic variance in the smaller central volumes where most of the observational data are, i.e. where they are the most effective.

In Figure \ref{fig:c2c}, at the starting redshift, the mean scatter of the random initial conditions is independent of the size of the sub-box with increasing variance in smaller sub-boxes. On the contrary the mean scatter of the 15 constrained initial conditions starts to decrease substantially below 300 \hMpc. As the median distance in {\it cosmicflows-2} is only 61 \hMpc\ and $\sim$ 98\% of the measurements are within 160 \hMpc, the standard deviation between density fields in the initial conditions is reduced in the expected large volume. The 10 constrained initial conditions with the same random realization field and added small scale features are on a line because adding modes at the scale of $\sim$ 0.98 (500/512) \hMpc\ does not affect the smoothed density distribution shown here. It only slightly increases the variance in the smallest sub-box.

The top of Figure \ref{fig:c2c2} shows the mean 1$\sigma$ scatters at redshift $z=0$ as a function of the sub-box size for the different types of simulations (random and constrained). The effect of the non-linear clustering is visible on small scales. As the system evolves and becomes more non-linear, a larger fraction of the sub-box is covered by low density regions (voids) than high density regions (clusters). Thus, voids become dominant in the volume. The densities of these nearly empty regions tend asymptotically to zero regardless of the initial field while that of high density regions, rising from small but positive differences in the initial field, are magnified. Consequently, the probability to compare a cell in a low density region with one in another low density region, with similar values, increases, reducing the scatter for each pair of random simulations. When considering sub-box smaller than 100 \hMpc, the 1$\sigma$ scatters decrease on average for the random simulations although the variance of these scatters increases because there is still the probability to find a high density region in one of the pair simulation versus a low density region in the other simulation of the pair. This is a limitation of the comparison method because the smaller the sub-box considered the higher the probability to find the same kind of field even if the probability to hit different types of fields is non null. The effect is not that pronounced for the constrained simulations because of the existence of similar (by construction) structures close to the center of the box (such as the Great Attractor and the Virgo cluster) and disappears for the constrained simulations with the same random realization field. We repeat the same procedure for the three components of the velocity field in the bottom panel of Figure \ref{fig:c2c2} and make the same observation. In addition, as a reference to assess the low values of the 1$\sigma$ scatters and thus the small discrepancies between the constrained simulated velocity fields, one can consider the validity of the Wiener-Filter reconstructed velocity field which is $\pm$ [100-150] \kms\ \citep{2015MNRAS.450.2644S}.

To summarize, Figures \ref{fig:c2c} and \ref{fig:c2c2} reveal that the cosmic variance, measured with the 1$\sigma$ scatter of cell-to-cell comparisons, is considerably reduced for constrained simulations, by a factor 2 to 3 on a scale of 5 \hMpc\ for both density and velocity fields in the inner part of the box, when compared to those obtained for random simulations. In addition, the error bars of these 1$\sigma$ scatters are smaller by a factor at least 2 when considering constrained simulations with respect to random simulations.  As an example, taking a sub-box of 150 \hMpc, the cosmic variance is decreased from 0.5 to 0.3 for the density fields normalized by the mean density at z=0 and from 150 to 50 \kms\ for the velocity fields. 

In Figure \ref{fig:c2c2}, we compare density and velocity fields, in (sub-)boxes, of constrained and random simulations. The observations, however, approximate a spherical distribution with a large radial dependence as demonstrated in Figure 1. Thus, corners of the (sub-)boxes appear at first to be less constrained than other regions at the edges of the (sub-)boxes. One might argue that spherical sub-volumes would be better suited for comparisons between random and constrained simulations. Using spherical regions to derive 1$\sigma$ scatters, we find no difference.

\begin{figure}
\flushleft
\hspace{-0.3cm}\includegraphics[width=0.48 \textwidth]{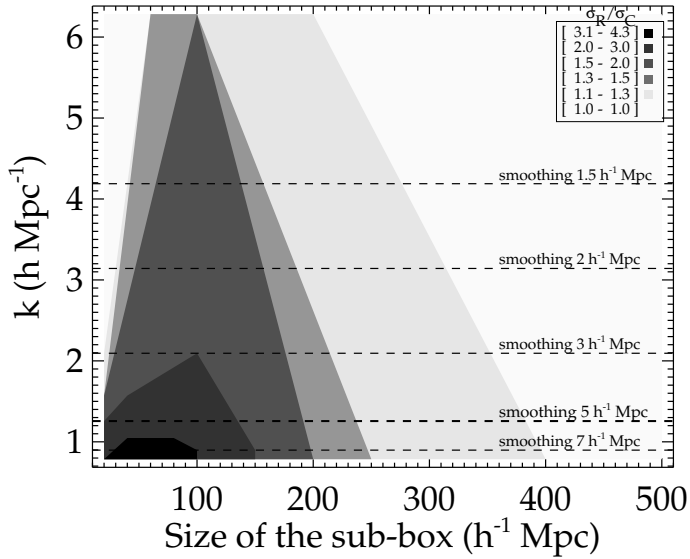}
\caption{constraining power, at z=0 of the dataset, combined with the method to build initial conditions, defined as the ratio of the mean one-sigma scatter obtained when comparing pairs of density fields normalized by the mean density of random simulations to that obtained for the pairs of density fields normalized by the mean density of constrained simulations. The higher the ratio, the darker the grey, the higher the constraining power (the smaller the cosmic variance), of the observational data combined with the method to build the initial conditions, is in a sub-box size versus resolution plane. The resolution is defined as $2\pi/R_s$ where $R_s$ is the value of the Gaussian smoothing in \hMpc.}
\label{fig:constrpower}
\end{figure}

Next, we apply different smoothing to the density fields normalized by the mean density to determine the constraining power, of the observational dataset combined with the method to build initial conditions, not only as a function of the box size but also as a function of the resolution. We define the resolution as $2\pi/R$ where $R$ is the value of the Gaussian smoothing in \hMpc, and the constraining power as the ratio of the mean 1$\sigma$ scatter obtained for the random simulations to that derived for the constrained simulations. We give to $R$ values ranging from 1 to 8 \hMpc. Figure \ref{fig:constrpower} display the constraining power of the observational peculiar velocities combined with the method to build initial conditions in a resolution versus sub-box size plane: the darker the grey, the higher the ratio of the 1$\sigma$ scatters, the higher the constraining power is (the more the cosmic variance is decreased in the constrained simulations with respect to the random simulations). As expected, the larger the smoothing (the smaller the resolution), the higher the constraining power is. On the other hand, with smaller smoothing (higher resolution), the cosmic variance is larger. In addition, the larger the sub-box size, the smaller the constraining power is because of the decreasing number of constraints. One can also notice that the effect of non-linearities becomes prominent as the smoothing decreases: small scales are essentially not constrained by the observational data. The non linear theory threshold is reached, on smaller scales shell crossing has wiped the initial correlations.

In the next section we compare the constrained simulations to the observed Local Universe.

\subsection{Simulation of the Large Scale Structure}

\begin{figure*}	
\vspace{-0.5cm}
\hspace{-1cm}\includegraphics[scale=1.1]{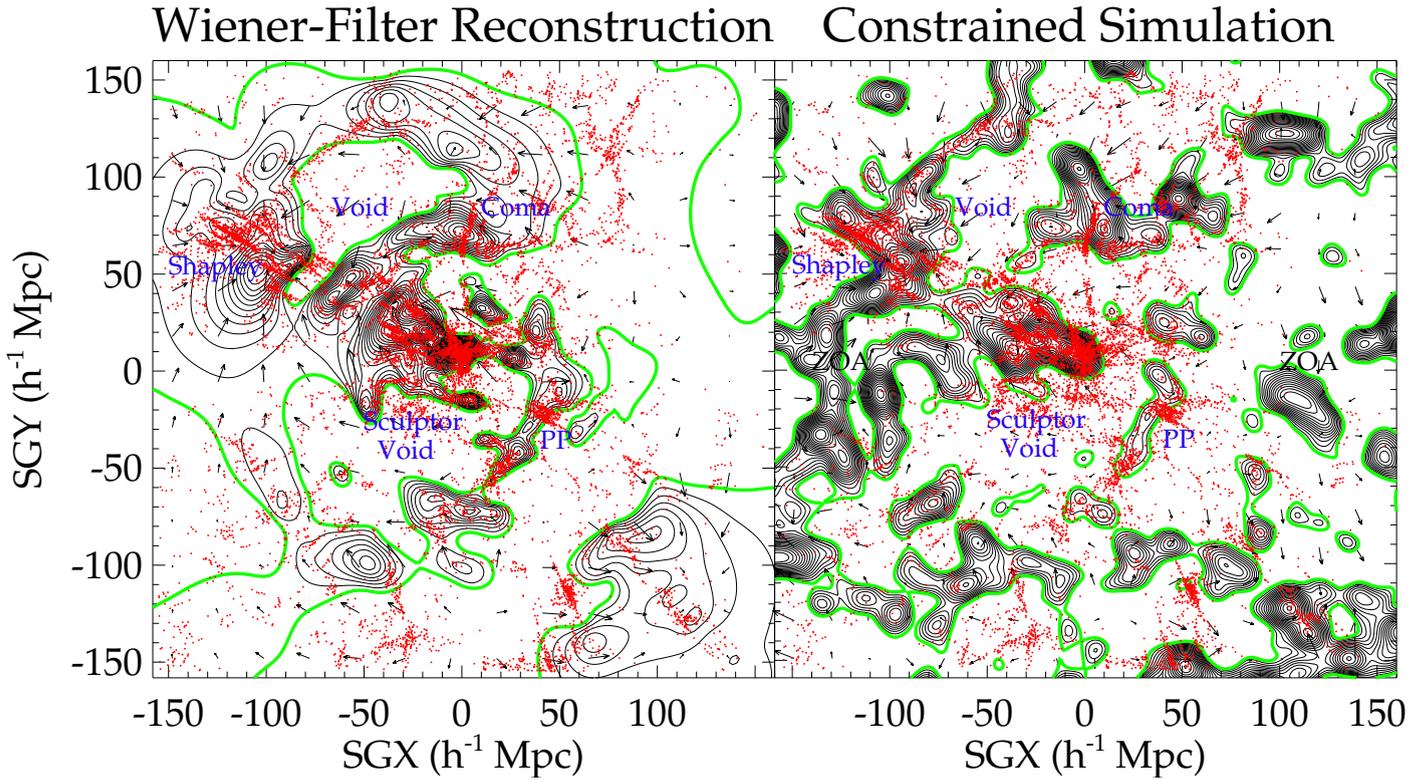}
\vspace{-0.06cm}
\caption{XY supergalactic plane of the reconstructed overdensity (contours) and velocity fields of the Local Universe obtained with the Wiener-Filter technique (left) and of the simulated density (contours) and velocity fields of one realization (right).The green color stands for the mean density. Arrows represent velocity fields. To facilitate the comparison, the simulation has been smoothed at 5 \hMpc\ and the reconstruction at 2 \hMpc\ which gives in both cases grid cells of $\sim$ 5 \hMpc. Galaxies from the 2MASS redshift catalog, in a $\pm$ 5 \hMpc\ thick slice, are superimposed as red dots for comparison purposes only. Fingers-of-god are visible in the galaxy distribution. Structures, voids and flows of the Local Universe are well recovered and simulated. A few of them are identified (blue names). While the Wiener-Filter reconstructs fairly well the Local Universe in the center of the box, the simulation allows to go farther in distances and deeper into the Zone of Avoidance (ZOA) and, more importantly, it supplies the whole density field (including non-linearities).}
\label{fig:WFsimuCF2}
\end{figure*}

In the previous subsection a study of the cosmic variance has demonstrated that the constrained simulations are remarkably similar to each other in the constrained part of the simulation box. To compare these simulations with the observed Local Universe, an observer is assumed to be at the center of the box and the three supergalactic coordinates are defined similarly to observational supergalactic coordinates. Comparing observations with simulations is not an easy task. Two possibilities are available although they both involved their own limitations: 1) comparing the Large Scale Structure in simulations to the distribution of observed galaxy surveys which are however magnitude limited, 2) comparing the fields of the simulations with the reconstructed ones obtained with the Wiener-Filter technique from the observational data. They constitute however only the linear fields and tend to the null fields in absence of data or in presence of noisy data. These limitations highlight the importance of the simulations which give access not only to the formation history but also to the full fields including non-linearities of the Local Universe.

Before comparing reconstructions, simulations and redshift surveys, we begin with a description of the observed structures in both the reconstruction and in one of the chosen randomly simulation as shown in Figure \ref{fig:WFsimuCF2}. Note that the choice of the simulation has no impact on the following discussion as simulations present the same Large Scale structure in the constrained $\sim$200 \hMpc\ radius area, namely high density regions and voids in this area are in every simulation. In this figure both the reconstructed and simulated (over)density (contour) and velocity (arrows) fields are displayed in a 5 \hMpc\ thick slice of the XY supergalactic plane. On top of the fields, red dots represent galaxies from the 2MASS redshift catalog in a 10 \hMpc\ thick slice. These galaxies are superimposed for comparisons purposes and one can notice fingers-of-god in the galaxy distribution. Several well-known structures can be identified in both the reconstructed and the simulated Local Universe: Perseus-Pisces (PP), Shapley as well as Coma superclusters but also voids such as the Sculptor void. In addition to the major structures and voids, the Zone of Avoidance (ZOA) due to our Milky-Way dust is marked. Note, that no structure have been reconstructed in that zone beyond 50 \hMpc\ from the center of the box due to a lack of information in the observed data. However, the simulation shows structures in this region, in particular connections between objects above and below the ZOA. 

Little is known about structures in the Zone of Avoidance but the simulation reproduces quite well the observations also in that zone:
\vspace{-0.2cm}
\begin{itemize}
\item A potential supercluster, at a distance about -60 \hMpc\ in the SGX direction, situated in the zone of obscuration \citep{1994ASPC...67...99K} is on an extension of the filament departing from Hydra and Antlia clusters going across the Zone Of Avoidance to reach the region of the Great Attractor \citep[$\sim$Centaurus Supercluster,][]{1994ASPC...67...99K}.
\item \citet{1994ASPC...67...99K} noted a clustering at a distance greater than -100 \hMpc\ in the SGX direction, in the zone hidden by our galaxy dust, a potential connection between the Horologium and Shapley Superclusters. The simulation contains a high density zone beyond -100 \hMpc\ in that direction. 
\end{itemize}

Regarding comparison with redshift surveys, there is qualitatively a good agreement between the 2MASS redshift catalog and the simulation on Figure \ref{fig:WFsimuCF2}. To assess this agreement we use the cosmic web based on the velocity shear tensor \citep{2012MNRAS.425.2049H} rather than on the gravitational tidal tensor \citep[e.g][]{2007MNRAS.381...41H} or on the displacement tensor \citep{2010MNRAS.403.1392L}. Eigenvalues of the velocity shear tensor permits to determine whether a region of the Universe with such a velocity field constitutes a knots, a filament, a sheet or a void. It is thus straightforward to determine the position of a galaxy in the cosmic web. With a null threshold and the definition used in \citet{2012MNRAS.425.2049H} for the velocity tensor, three negative eigenvalues correspond to a void while three positive value correspond to a knot. Two negative and one positive value constitute a sheet while the opposite stands for a filament. If simulations are in good agreement with observations, there should be approximately the same number of galaxies in filament and sheets  ($\sim$ 35-45 \%) then less in knots and in voids ($\sim$ 10 \%) \citep[e.g.][although they choose a slightly higher than zero threshold, while we choose zero, they checked that general results are quasi independent of the threshold choice as long as the chosen value is reasonable]{2015ApJ...799...45F,2012MNRAS.421L.137L}. On average of the fifteen cosmic webs computed from the fifteen different constrained simulations, 6$\pm$1 \% of the galaxies are in knots, 35$\pm$2 \% are in filaments, 48$\pm$2 \% are in sheets and 10$\pm$1 \% are in voids. The galaxies are distributed as expected giving agreement with observations.

\begin{figure*}
\includegraphics[width=0.46 \textwidth]{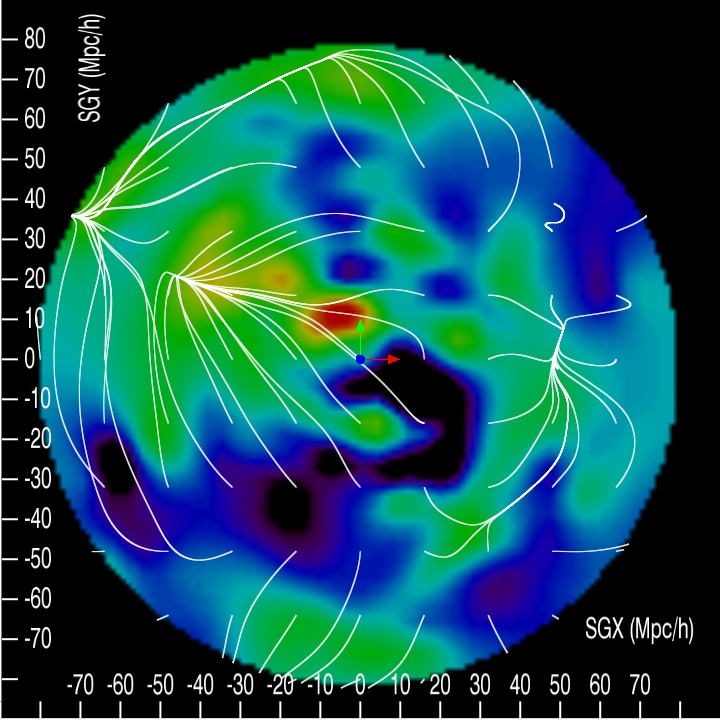}
\includegraphics[width=0.46 \textwidth]{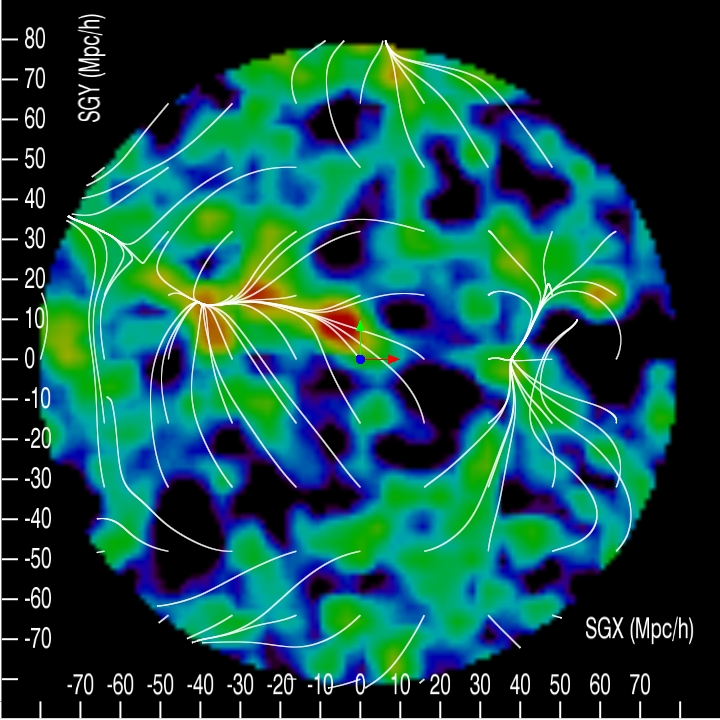}
\caption{Supergalactic XY plane of the reconstruction (left) and of one constrained simulation (right) of the Local Universe. The gradient of colors stand for the overdensity (reconstruction) and density (simulation) fields . Voids are in blue-black and high density regions are in yellow-red. The divergent velocity field due to densities within 100 \hMpc\ around the Laniakea Supercluster center at [-47,13,-5] \hMpc\ is represented by streamlines. Reconstructed and simulated streamlines look alike.}
\label{fig:Laniakea}
\end{figure*}

As for comparisons with the reconstruction, coming back to Figure \ref{fig:WFsimuCF2}, the Wiener-Filter reconstructs fairly well the Local Universe in the center of the box although it shows only the linear fields and tends to the null field in absence of data or in presence of noisy data, thus reconstruction and simulation agree qualitatively very well: the reconstruction presents more feature in the center relatively to its edges but the loss of precision with the distance from the center of the box is the cause. The simulation allows one to go deeper into the Zone of Avoidance and to extend further the study of the Large Scale Structure and, more importantly, it supplies the whole density field, including non-linearities. An estimation of the agreement between reconstruction and simulations can be made with cell-to-cell comparisons between the velocity fields which, unlike densities, are highly linear. The 1$\sigma$ scatter is on average of the order of 100-150 \kms\ (i.e. 2-3 \hMpc, the linear theory threshold, in terms of displacement). From this comparison between a simulation and the reconstruction, it can be concluded that the major attractors and voids of the Local Universe are properly simulated.

The variance between the different constrained simulations is relatively low (see previous subsection). If a Large Scale Structure feature is present in one of the simulations, there is a high probability to recover it in all the other realizations. The Large Scale Environment is robustly simulated. Since this Large Scale environment has been suggested to play an essential role in the formation and evolution of local objects \citep[e.g.][]{2014MNRAS.438.2578G}, these constrained simulations are ideal to study local objects. First, we turn towards a recently discovered Large Scale feature in our neighborhood, the Laniakea supercluster of galaxies \citep{2014Natur.513...71T}.

\subsection{Example of the Laniakea Supercluster}

In this subsection, we focus on a particular structure of the Local Universe, the Laniakea Supercluster of galaxies discovered and defined in \citet{2014Natur.513...71T}. This supercluster is constituted of a local basin of attraction and every objects with a perturbative motion toward it. Local flows encompassed in that region converge onto the local attractor. To compare the simulated superclusters with the observed-reconstructed one, the divergent field is evaluated. For a chosen volume, the divergent field corresponds to velocities due solely to densities in this volume. In order to remove most of the non linear components, the density and velocity grids are smoothed with a 5\hMpc\ Gaussian. With grids obtained via a cloud-in-cell scheme and a 60 \hMpc\ sphere radius centered at [-47,13,-5] \hMpc\ similarly to the definition given by \citet{2014Natur.513...71T}, we are able to find the local basin of attraction in the different simulations. Enlarging the radius, the contours of the simulated Laniakea superclusters of galaxies are recovered. The result obtained with one realization is given on the right panel of Figure \ref{fig:Laniakea}. The left panel shows the Wiener-Filter reconstructed supercluster. In this figure, the gradient of colors correspond to densities while streamlines stand for the velocity fields. From blue to red the density increases, namely voids are in blue-black and high density regions are in yellow-red. The simulation and the reconstruction look alike. The Laniakea supercluster streamlines are very well simulated and converge in a similar location about [$-47$,11,0] \hMpc. The Laniakea supercluster is surrounded by cosmic gravitational streams flowing towards the Perseus-Pisces superclusters on the positive SGX side, towards Shapley on the negative SGX side and towards Coma on the positive SGY side in both the simulation and the reconstruction. \citet{2014Natur.513...71T} estimate the mass of the supercluster at around 6.5 $\times$ 10$^{16}$ $\hmsun$ on the scale of this paper, in relatively good agreement with simulations where the number of dark matter particles contained in the Laniakea supercluster, simplified as a 60 \hMpc\ sphere radius centered at [-47,13,-5] \hMpc, gives a mass of approximately 2 $\pm$ 0.3 $\times$ 10$^{16}$ $\hmsun$ over the different realizations.

To evaluate the agreement between the Wiener-Filter reconstruction and the simulations, we proceed as in the previous section with cell-to-cell comparisons within the Laniakea supercluster region. The 1$\sigma$ scatter is 104 \kms\ on average with a standard deviation of 4 \kms, the median is identical to the mean. Cell-to-cell comparisons in this region between constrained simulations give on average 1$\sigma$ scatters of 45 $\pm$ 6 \kms\ for the velocity fields and of 0.29 $\pm$ 0.02 for the density fields at z=0. By comparison, the average 1$\sigma$ scatters obtained when comparing random simulations in that region are about 145 $\pm$ 35 \kms\ for the velocity fields and of 0.43 $\pm$ 0.09 for the density fields.

\subsection{Observed Clusters \& Simulated Dark Matter Halos}

Finally, we turn our attention to the study of halos at redshift zero. Lists of dark matter halos obtained with the Amiga Halo Finder \citep{2009ApJS..182..608K} are used to match halos in simulations with clusters in the Local Universe. Virgo is one of the largest cluster in our neighborhood and it is the closest. It is natural to look for candidates of that cluster as it should be the most efficiently constrained. In addition, attempts to find candidates for Centaurus (restricted to one component of a complex in the region), Coma and Perseus are made.

\begin{figure}
\vspace{-0cm}
\includegraphics[width=0.45 \textwidth]{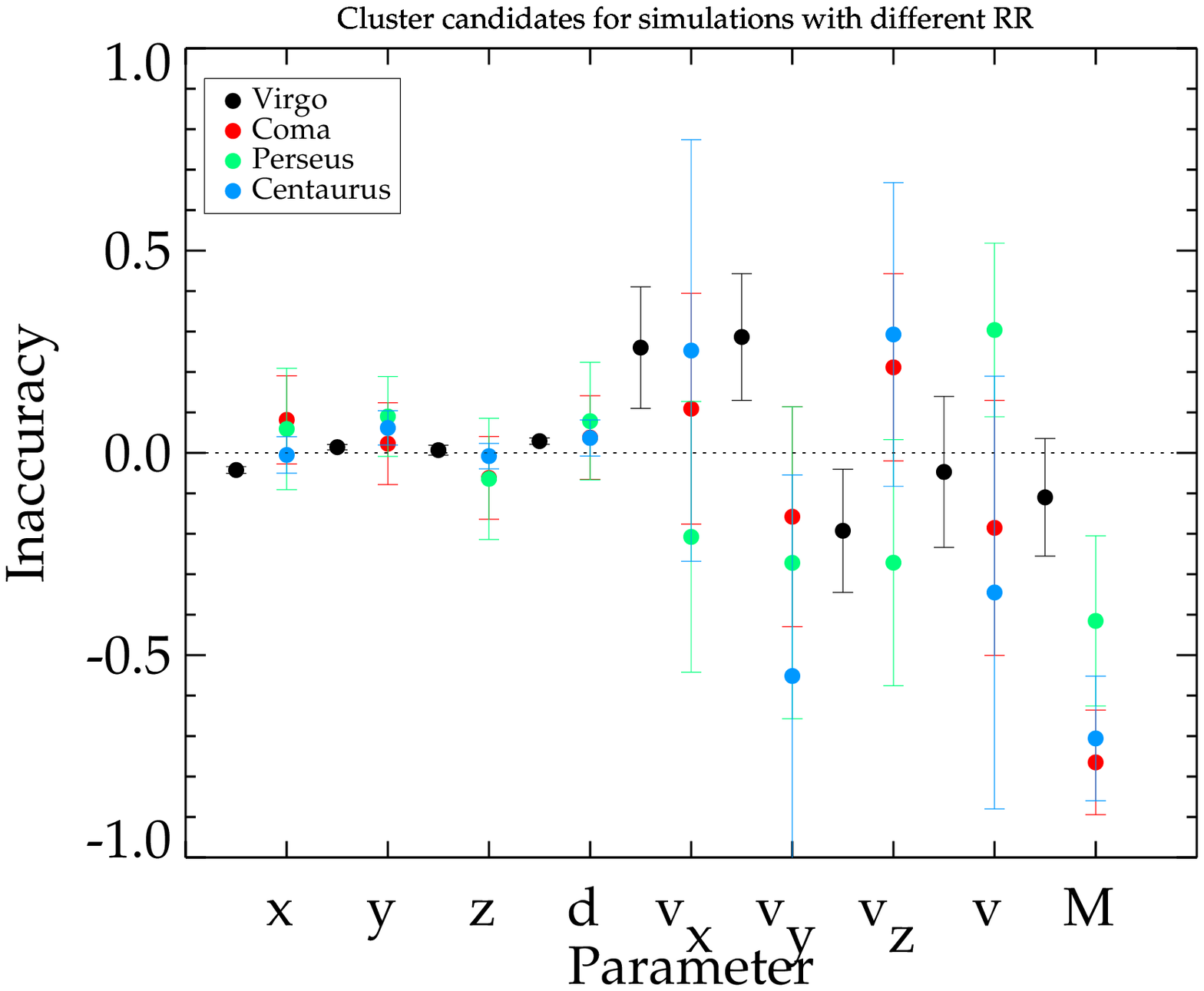}\\

\vspace{-0cm}
\includegraphics[width=0.45 \textwidth]{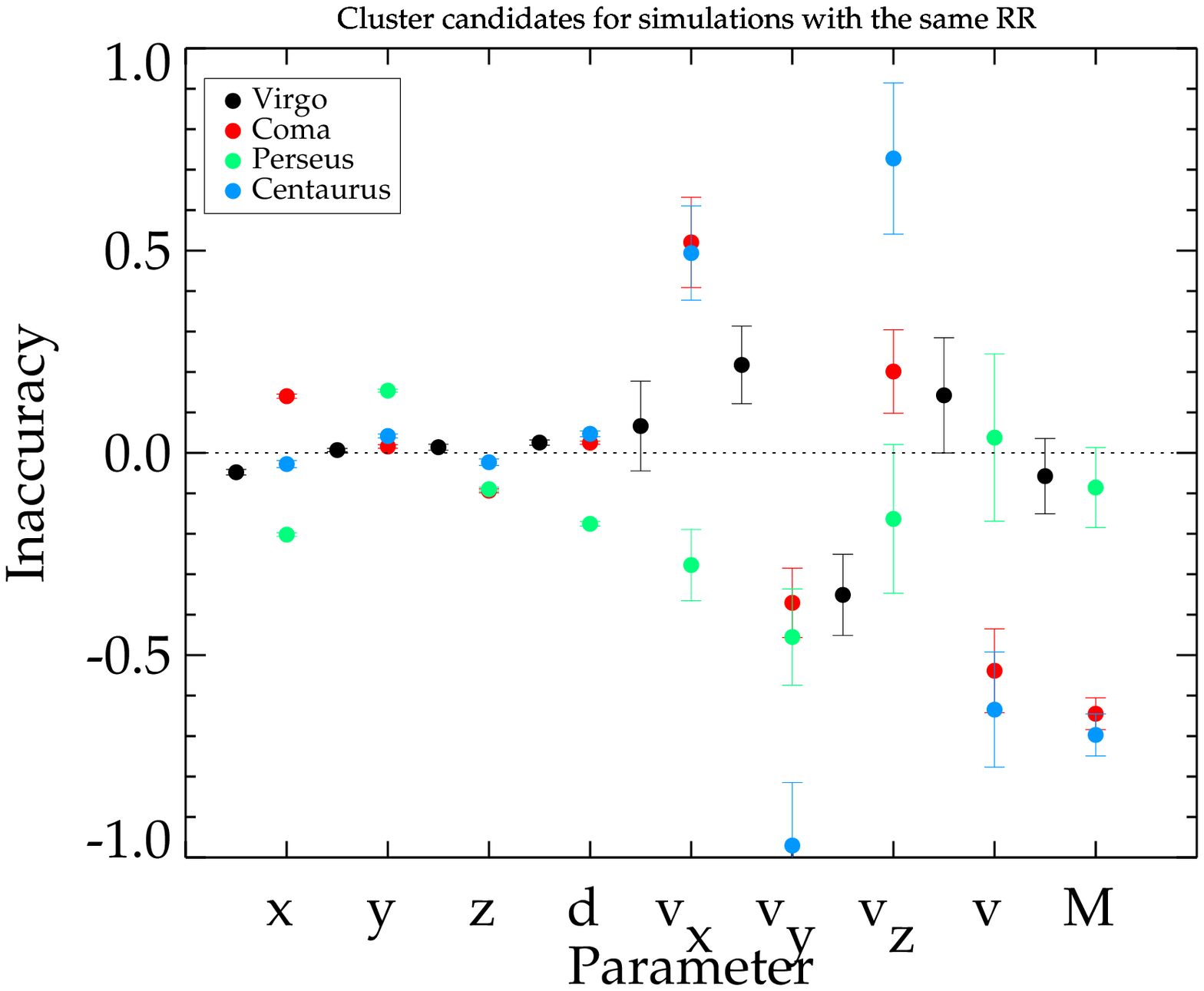}\\

\vspace{-0.5cm}
\caption{Inaccuracies of the parameters of Virgo, Coma, Perseus and Centaurus dark matter halo candidates. The inaccuracy is defined as the difference between the simulated and the observed or Wiener-Filter reconstructed component divided by respectively the mean distance in {\it cosmicflows-2} for the supergalactic coordinates and the distance, the mean velocity of halos of approximately the same mass as candidates for the velocity components and the velocity vector and the mean estimated mass of the observed cluster for the mass. Top: Candidates in the fifteen constrained simulations based on different random realization fields. Bottom: Candidates in the ten constrained simulations sharing the same random realization field. For a clearer visibility, datapoints corresponding to Virgo candidates are shifted to the left on the x-axis with respect to datapoints for other cluster candidates. As expected, Virgo candidates are the best constrained because they are close to the center of the box.}
\label{fig:Virgoinaccuracy}
\end{figure}

We begin with the Virgo cluster. Candidates are found at less than 3-4 \hMpc\ from its observational position in all the simulations. The replicas have masses (M$_{200}$ with respect to the critical density) for the simulations between 2.7 and 4.3 $\times$ 10$^{14}$ h$^{-1}$$M_\odot$ which is in good agreement with current values given for this cluster \citep[e.g. 2 - 6  $\times$ 10$^{14}$ $\hmsun$ in Planck cosmology,][]{2010MNRAS.405.1075K}, especially considering that within simulations, masses of cluster candidates can vary between half and twice the cluster mass \citep{2011MNRAS.413.1961L} and that the estimated mass of a cluster depends on the method/definition used, as does the mass of a simulated halo. Several parameters can be compared between cluster candidates and the observed cluster. To estimate the agreement between candidates and observed Virgo cluster, we define the inaccuracy as the difference between the simulated and the observed (mass, distance, coordinate) or Wiener-Filter reconstructed (velocity) component divided by: 1) the mean distance in {\it cosmicflows-2} for the coordinates and distances, 2) the mean velocity of halos of approximately the same mass as Virgo candidates for the velocities and their components, 3) the mean estimated mass of the observed clusters for the masses. Note that the definition we choose for the inaccuracy differs from that of the relative change or difference as we normalize neither by the reference (i.e. observed or reconstructed) parameter nor by a function of the simulated and observed or reconstructed parameters. We justify this choice by the fact that further-from-the-box-center halos would artificially appear on Figure \ref{fig:Virgoinaccuracy} in better agreement with the observed clusters than their counterparts using the latter definitions (i.e. division by the distance). The same artificial observation would happen if more massive (faster) halos were normalized by the mass (velocity) of the halos. For this reason, the normalization is made with reference values independent of the cluster characteristics. The compared parameters are the three supergalactic coordinates and the square root of the sum of their squares, namely the distance, the three components of the velocity and the 3D velocity and the mass (M$_{200}$ with respect to the critical density for the simulations).

Figure \ref{fig:Virgoinaccuracy} gathers the mean and scatter of the inaccuracies of the parameters as defined above for the Virgo candidates found in the different constrained simulations. The top panel shows the mean inaccuracies (filled black circles) and their standard deviations (error bars) for the Virgo candidates in the 15 simulations with a different random realization field while the bottom panel gives inaccuracies and standard deviations of Virgo candidates found in the ten simulations sharing the same random realization field. There are two observations: 1) there is a very good agreement between the observed-reconstructed and simulated parameters as inaccuracies are close to zero. 2) the standard deviation between the different simulations are quite small (less than 10-15\% of the considered parameter). As expected, the ten simulations which differ only on the random small scale features give Virgo candidates in slightly better agreement with each other (scatter less than 5-10\% of the considered parameter).

We proceed similarly for Centaurus, Coma and Perseus although we expect candidates to be somewhat more scattered because the simulations, although constrained on the Large Scale, are not as well constrained as they are in the very center of the box, near Virgo candidates. Rather than using the estimated masses, we settle for looking for halos more massive than 5 10$^{13}$ $\hmsun$ close to the positions of the observed clusters. If present these halos confirm a high density regions where expected. In only one of the fifteen constrained simulations, we cannot locate a halo massive enough in a reasonable radius around the observed position of Coma and Centaurus clusters. As for the Centaurus location, in two simulations we have two candidates of approximately the same mass nearby (the distance between the two is less than $\approx$3 \hMpc), we select the closest to the estimated observed position. Note that finding two halos close to each other is not surprising as Centaurus is constituted of several massive components. We then proceed as for Virgo candidates, i.e. we derive the inaccuracies for Centaurus, Coma and Perseus candidates although we fix the mass estimate to 5 $\times$ 10$^{14}$ $\hmsun$ as the reference for these three other clusters. We justify our choice by the fact that 1) simulations are not as well constrained at the positions of these other clusters as they are at Virgo's position, 2) the estimated mass of these clusters varies with the measurement method and the defined boundaries of the cluster (as an example, the estimated mass of the Coma cluster is 5.1 $\times$ 10$^{14}$ $\hmsun$ according to weak lensing measurements in \citet{2009A&A...498L..33G} while it is 1.4 $\times$ 10$^{15}$ $\hmsun$ when using the radius of second turn around as define by \cite{2010arXiv1010.3787T}) and 3) the definition of the observed mass is different from the mass of simulated halos. Results are given in Figure \ref{fig:Virgoinaccuracy} with three different colors for the three different clusters. They are slightly shifted to the right on the x-axis with respect to datapoints obtained for Virgo candidates for a clearer visibility. As for Virgo candidates, inaccuracies are small in terms of positions in agreement with the fact that the Large Scale Structure is well constrained. Scatters are larger than for the Virgo candidates but that was expected as we are looking at less constrained regions. This is not surprising that the higher scatter in velocities is observed for the Centaurus-like halos as Centaurus is an ensemble of objects rather than a compact object: non-linear motions are involved in that zone and because these regions are very dense, there are different massive halos with a high probability to pick a different one in each one of the realization. Looking at the bottom of Figure \ref{fig:Virgoinaccuracy} which shows candidates found in the simulations sharing the same random realization field, the scatters are decreased by at least a factor 2.  

Only objects with low mass are heavily modified by the random small scale features added to increase the resolution while the Large Scale Structure is quite unaffected by them. Considering that a goal of the CLUES project is to build a large statistical sample of Local Group-like entities, selecting simulations containing all the appropriate clusters-like objects, we will be able to build a factory of look-alikes of the Local Group in the proper environment to study their statistical properties due to both the Large Scale Structure (constrained with different random realizations) and to the small scale structures (constrained but sharing the same random seed) in a decoupled way.


\section{Conclusion}

The first generation of simulations constrained by observational peculiar velocities produced by the CLUES project was affected by a substantial shift in the positions of objects recovered at redshift zero. In this paper, we present a double improvement with respect to this first generation. First, we use the second catalog of the observational Cosmicflows project which is superior in size (number of constraints), extent ($\sim$ up to 150 \hMpc) and accuracy compared to the previously used catalogs. Second, we add the newly developed techniques involving the grouping of galaxies, the minimization of biases, and the reverse Zel'dovich approximation based on the Wiener-Filter method, with the constrained realization technique to build more accurate constrained initial conditions. We are able to show that not only constrained simulations exhibit a lower cosmic variance than random simulations but also that they are in agreement with our cosmic neighborhood up to the non-linear scale (2-3 \hMpc). \\
To do so, we compare a set of 15 random and 25 constrained simulations (10 of the 25 simulations share the same Large Scale random phase and differ only by the small waves added to increase the resolution) of 512$^3$ particles within a 500 \hMpc\ boxsize. A check with two 1024$^3$-particles simulations showed that the results are not affected by the number of particles. We apply a cloud-in-cell scheme to all the simulations and smooth the resulting velocity and density grids with a 5 \hMpc\ Gaussian. \\
We define the cosmic variance as the one-sigma, scatter (or standard deviation) in density-density plots (the field of a first simulation versus the field of a second simulation) obtained from cell-to-cell comparisons between pairs of simulations of the same nature (random or constrained). We average the results over the different pairs and found that the 1$\sigma$ scatters obtained for constrained simulations are not only minimal when comparing the inner part of the boxes, where most of the constraints are, but they are also smaller by a factor 2 to 3 with respect to those found for random simulations. The best constrained part of the simulations is the inner box within approximately 100 \hMpc\ for the smallest (clusters) scales, the resemblance extends to 300 \hMpc\ on larger scales (5 to a few tens of megaparsecs). This agreement meets expectations as the {\it cosmicflows-2} catalog extends to 230 \hMpc\ with 98\% of the distance measurements within 160 \hMpc\ and a median distance of 61 \hMpc. \\
We found that on average the constrained simulations tend to have less power on large scales than the random simulations, although they are well within the expected scatter.  This effect could be due to the observational data and/or to the way they are processed to build the initial conditions. We are working on an improved algorithm which reduces or removes this effect by taking into consideration the reduced accuracy of the reconstruction on large distances. This will be important when studying large scale velocity flows. However, results presented in this paper are in no way affected by this effect. In fact, adding artificially power on the largest modes would only slightly change the mass of the most massive objects. We will study and discuss this effect in more detail with a series of new simulations and discuss it in a forthcoming paper.\\
To compare the simulations with the observed Local Universe, we use cell-to-cell comparisons between the reconstructed and the simulated velocity field. We find that simulations at redshift zero agree with the Wiener-Filter reconstruction obtained with the observations at 100-150 \kms\ or 2-3 \hMpc, namely the linear theory threshold. \\
Taking as an example the Laniakea Supercluster of galaxies, defined as a local basin of attraction and all flows going towards it, we show that simulated and reconstructed Laniakea superclusters are in relatively good agreement. The mean 1$\sigma$ scatter obtained from cell-to-cell comparisons between the reconstructed and simulated velocity fields is 104 $\pm$ 4 \kms. When comparing the simulations, the mean 1$\sigma$ scatter of the simulated Laniakea superclusters' fields is 45 $\pm$ 6 \kms\ for their velocity fields and 0.29 $\pm$ 0.02 for their density fields. By comparison, similar regions between random simulations differ by 145 $\pm$ 35 \kms\ for the velocity fields and 0.43 $\pm$ 0.09 for the density fields. \\
Finally, we give an overview of Virgo candidates as well as other well-known nearby clusters at redshift zero and show that again the scatter between simulated dark matter halo candidates within themselves and also with the observed-reconstructed clusters in terms of position, velocity and mass is only of the order of 10\%.\\
These comparisons show that simulations are in agreement between each other and above all with the reconstruction. Because the reconstruction recovers fairly well all the major attractors and voids of the Local Universe, they must also be present in all the simulations at redshift zero for these latter to be similar to the reconstruction, i.e. to the observations. The method to build more accurate constrained initial conditions is extremely efficient. We produced the first simulations constrained with observational radial peculiar velocities which resemble the Local Universe up to 150 \hMpc, with increasing accuracy when reaching the inner part of the box.

\section*{Acknowledgements}  
The simulations have been performed at the Leibniz Rechenzentrum (LRZ) in Munich and at J\"ulich Supercomputing Centre. We thank the anonymous referee whose comments helped improving the manuscript. JS acknowledges support from the Alexander von Humboldt Foundation. SG and YH acknowledge support from DFG under the grant GO563/21-1. GY acknowledges support from the Spanish MINECO under research grants AYA2012-31101 and  FPA2012-34694. YH has been partially supported by the Israel Science Foundation (1013/12). HC acknowledges support from the Lyon Institute of Origins under grant ANR-10-LABX-66 and from CNRS under PICS-06233. RBT received support from the NASA Astrophysics Data Analysis Program.


\bibliographystyle{mnras}

\bibliography{biblicomplete}

\begin{thebibliography}{80}
\expandafter\ifx\csname natexlab\endcsname\relax\def\natexlab#1{#1}\fi

\bibitem[{{Abazajian} {et~al}\mbox{.}(2003){Abazajian}, {Adelman-McCarthy},
  {Ag{\"u}eros}, {Allam}, {Anderson}, {Annis}, {Bahcall}, {Baldry}, {Bastian},
  {Berlind}, {Bernardi}, {Blanton}, {Blythe}, {Bochanski}, {Boroski},
  {Brewington}, {Briggs}, {Brinkmann}, {Brunner}, {Budav{\'a}ri}, {Carey},
  {Carr}, {Castander}, {Chiu}, {Collinge}, {Connolly}, {Covey}, {Csabai},
  {Dalcanton}, {Dodelson}, {Doi}, {Dong}, {Eisenstein}, {Evans}, {Fan},
  {Feldman}, {Finkbeiner}, {Friedman}, {Frieman}, {Fukugita}, {Gal},
  {Gillespie}, {Glazebrook}, {Gonzalez}, {Gray}, {Grebel}, {Grodnicki}, {Gunn},
  {Gurbani}, {Hall}, {Hao}, {Harbeck}, {Harris}, {Harris}, {Harvanek},
  {Hawley}, {Heckman}, {Helmboldt}, {Hendry}, {Hennessy}, {Hindsley}, {Hogg},
  {Holmgren}, {Holtzman}, {Homer}, {Hui}, {Ichikawa}, {Ichikawa}, {Inkmann},
  {Ivezi{\'c}}, {Jester}, {Johnston}, {Jordan}, {Jordan}, {Jorgensen},
  {Juri{\'c}}, {Kauffmann}, {Kent}, {Kleinman}, {Knapp}, {Kniazev}, {Kron},
  {Krzesi{\'n}ski}, {Kunszt}, {Kuropatkin}, {Lamb}, {Lampeitl}, {Laubscher},
  {Lee}, {Leger}, {Li}, {Lidz}, {Lin}, {Loh}, {Long}, {Loveday}, {Lupton},
  {Malik}, {Margon}, {McGehee}, {McKay}, {Meiksin}, {Miknaitis}, {Moorthy},
  {Munn}, {Murphy}, {Nakajima}, {Narayanan}, {Nash}, {Neilsen}, {Newberg},
  {Newman}, {Nichol}, {Nicinski}, {Nieto-Santisteban}, {Nitta}, {Odenkirchen},
  {Okamura}, {Ostriker}, {Owen}, {Padmanabhan}, {Peoples}, {Pier}, {Pindor},
  {Pope}, {Quinn}, {Rafikov}, {Raymond}, {Richards}, {Richmond}, {Rix},
  {Rockosi}, {Schaye}, {Schlegel}, {Schneider}, {Schroeder}, {Scranton},
  {Sekiguchi}, {Seljak}, {Sergey}, {Sesar}, {Sheldon}, {Shimasaku}, {Siegmund},
  {Silvestri}, {Sinisgalli}, {Sirko}, {Smith}, {Smol{\v c}i{\'c}}, {Snedden},
  {Stebbins}, {Steinhardt}, {Stinson}, {Stoughton}, {Strateva}, {Strauss},
  {SubbaRao}, {Szalay}, {Szapudi}, {Szkody}, {Tasca}, {Tegmark}, {Thakar},
  {Tremonti}, {Tucker}, {Uomoto}, {Vanden Berk}, {Vandenberg}, {Vogeley},
  {Voges}, {Vogt}, {Walkowicz}, {Weinberg}, {West}, {White}, {Wilhite},
  {Willman}, {Xu}, {Yanny}, {Yarger}, {Yasuda}, {Yip}, {Yocum}, {York},
  {Zakamska}, {Zehavi}, {Zheng}, {Zibetti}, \& {Zucker}}]{2003AJ....126.2081A}
{Abazajian} K. {et~al.}, 2003, \aj, 126, 2081

\bibitem[{{Abazajian} {et~al}\mbox{.}(2009){Abazajian}, {Adelman-McCarthy},
  {Ag{\"u}eros}, {Allam}, {Allende Prieto}, {An}, {Anderson}, {Anderson},
  {Annis}, {Bahcall}, \& et~al.}]{2009ApJS..182..543A}
{Abazajian} K.~N. {et~al.}, 2009, \apjs, 182, 543

\bibitem[{{Alimi} {et~al}\mbox{.}(2012){Alimi}, {Bouillot}, {Rasera},
  {Reverdy}, {Corasaniti}, {Balmes}, {Requena}, {Delaruelle}, \&
  {Richet}}]{DeusSimulation2012}
{Alimi} J.-M. {et~al.}, 2012, ArXiv e-prints: 1206.2838

\bibitem[{{Angulo} {et~al}\mbox{.}(2012){Angulo}, {Springel}, {White},
  {Jenkins}, {Baugh}, \& {Frenk}}]{AnguloXXL2012}
{Angulo} R.~E., {Springel} V., {White} S.~D.~M., {Jenkins} A., {Baugh} C.~M.,
  {Frenk} C.~S., 2012, \mnras, 426, 2046

\bibitem[{{Bertschinger} {et~al}\mbox{.}(1990){Bertschinger}, {Dekel}, {Faber},
  {Dressler}, \& {Burstein}}]{1990ApJ...364..370B}
{Bertschinger} E., {Dekel} A., {Faber} S.~M., {Dressler} A., {Burstein} D.,
  1990, \apj, 364, 370

\bibitem[{{Colless} {et~al}\mbox{.}(2001){Colless}, {Saglia}, {Burstein},
  {Davies}, {McMahan}, \& {Wegner}}]{2001MNRAS.321..277C}
{Colless} M., {Saglia} R.~P., {Burstein} D., {Davies} R.~L., {McMahan} R.~K.,
  {Wegner} G., 2001, \mnras, 321, 277

\bibitem[{{Dekel}(1994)}]{1994ARA&A..32..371D}
{Dekel} A., 1994, \araa, 32, 371

\bibitem[{{Dekel} {et~al}\mbox{.}(1990){Dekel}, {Bertschinger}, \&
  {Faber}}]{1990ApJ...364..349D}
{Dekel} A., {Bertschinger} E., {Faber} S.~M., 1990, \apj, 364, 349

\bibitem[{{Doumler} {et~al}\mbox{.}(2013{\natexlab{a}}){Doumler}, {Courtois},
  {Gottl{\"o}ber}, \& {Hoffman}}]{2013MNRAS.430..902D}
{Doumler} T., {Courtois} H., {Gottl{\"o}ber} S., {Hoffman} Y.,
  2013{\natexlab{a}}, \mnras, 430, 902

\bibitem[{{Doumler} {et~al}\mbox{.}(2013{\natexlab{b}}){Doumler},
  {Gottl{\"o}ber}, {Hoffman}, \& {Courtois}}]{2013MNRAS.430..912D}
{Doumler} T., {Gottl{\"o}ber} S., {Hoffman} Y., {Courtois} H.,
  2013{\natexlab{b}}, \mnras, 430, 912

\bibitem[{{Doumler} {et~al}\mbox{.}(2013{\natexlab{c}}){Doumler}, {Hoffman},
  {Courtois}, \& {Gottl{\"o}ber}}]{2013MNRAS.430..888D}
{Doumler} T., {Hoffman} Y., {Courtois} H., {Gottl{\"o}ber} S.,
  2013{\natexlab{c}}, \mnras, 430, 888

\bibitem[{{Forero-Romero} \& {Gonz{\'a}lez}(2015)}]{2015ApJ...799...45F}
{Forero-Romero} J.~E., {Gonz{\'a}lez} R., 2015, \apj, 799, 45

\bibitem[{{Freedman} {et~al}\mbox{.}(2001){Freedman}, {Madore}, {Gibson},
  {Ferrarese}, {Kelson}, {Sakai}, {Mould}, {Kennicutt}, {Ford}, {Graham},
  {Huchra}, {Hughes}, {Illingworth}, {Macri}, \&
  {Stetson}}]{2001ApJ...553...47F}
{Freedman} W.~L. {et~al.}, 2001, \apj, 553, 47

\bibitem[{{Ganon} \& {Hoffman}(1993)}]{1993ApJ...415L...5G}
{Ganon} G., {Hoffman} Y., 1993, \apjl, 415, L5

\bibitem[{{Garrison-Kimmel} {et~al}\mbox{.}(2014){Garrison-Kimmel},
  {Boylan-Kolchin}, {Bullock}, \& {Lee}}]{2014MNRAS.438.2578G}
{Garrison-Kimmel} S., {Boylan-Kolchin} M., {Bullock} J.~S., {Lee} K., 2014,
  \mnras, 438, 2578

\bibitem[{{Gavazzi} {et~al}\mbox{.}(2009){Gavazzi}, {Adami}, {Durret},
  {Cuillandre}, {Ilbert}, {Mazure}, {Pell{\'o}}, \&
  {Ulmer}}]{2009A&A...498L..33G}
{Gavazzi} R., {Adami} C., {Durret} F., {Cuillandre} J.-C., {Ilbert} O.,
  {Mazure} A., {Pell{\'o}} R., {Ulmer} M.~P., 2009, \aap, 498, L33

\bibitem[{{Gottl\"ober} {et~al}\mbox{.}(2010){Gottl\"ober}, {Hoffman}, \&
  {Yepes}}]{2010arXiv1005.2687G}
{Gottl\"ober} S., {Hoffman} Y., {Yepes} G., 2010, ArXiv e-prints: 1005.2687

\bibitem[{{Hahn} {et~al}\mbox{.}(2007){Hahn}, {Carollo}, {Porciani}, \&
  {Dekel}}]{2007MNRAS.381...41H}
{Hahn} O., {Carollo} C.~M., {Porciani} C., {Dekel} A., 2007, \mnras, 381, 41

\bibitem[{{Han}(1992)}]{1992ApJ...395...75H}
{Han} M., 1992, \apj, 395, 75

\bibitem[{{Hendry} \& {Simmons}(1994)}]{1994ApJ...435..515H}
{Hendry} M.~A., {Simmons} J.~F.~L., 1994, \apj, 435, 515

\bibitem[{{He{\ss}} {et~al}\mbox{.}(2013){He{\ss}}, {Kitaura}, \&
  {Gottl{\"o}ber}}]{2013MNRAS.435.2065H}
{He{\ss}} S., {Kitaura} F.-S., {Gottl{\"o}ber} S., 2013, \mnras, 435, 2065

\bibitem[{{Hoffman} {et~al}\mbox{.}(2012){Hoffman}, {Metuki}, {Yepes},
  {Gottl{\"o}ber}, {Forero-Romero}, {Libeskind}, \&
  {Knebe}}]{2012MNRAS.425.2049H}
{Hoffman} Y., {Metuki} O., {Yepes} G., {Gottl{\"o}ber} S., {Forero-Romero}
  J.~E., {Libeskind} N.~I., {Knebe} A., 2012, \mnras, 425, 2049

\bibitem[{{Hoffman} \& {Ribak}(1991)}]{1991ApJ...380L...5H}
{Hoffman} Y., {Ribak} E., 1991, \apjl, 380, L5

\bibitem[{{Hoffman} \& {Ribak}(1992)}]{1992ApJ...384..448H}
{Hoffman} Y., {Ribak} E., 1992, \apj, 384, 448

\bibitem[{{Huchra} {et~al}\mbox{.}(2012){Huchra}, {Macri}, {Masters},
  {Jarrett}, {Berlind}, {Calkins}, {Crook}, {Cutri}, {Erdo{\v g}du}, {Falco},
  {George}, {Hutcheson}, {Lahav}, {Mader}, {Mink}, {Martimbeau}, {Schneider},
  {Skrutskie}, {Tokarz}, \& {Westover}}]{2012ApJS..199...26H}
{Huchra} J.~P. {et~al.}, 2012, \apjs, 199, 26

\bibitem[{{Hudson}(1994)}]{1994MNRAS.266..468H}
{Hudson} M.~J., 1994, \mnras, 266, 468

\bibitem[{{Jasche} \& {Wandelt}(2013)}]{2013MNRAS.432..894J}
{Jasche} J., {Wandelt} B.~D., 2013, \mnras, 432, 894

\bibitem[{{Jha} {et~al}\mbox{.}(2007){Jha}, {Riess}, \&
  {Kirshner}}]{2007ApJ...659..122J}
{Jha} S., {Riess} A.~G., {Kirshner} R.~P., 2007, \apj, 659, 122

\bibitem[{{Karachentsev} {et~al}\mbox{.}(2004){Karachentsev}, {Karachentseva},
  {Huchtmeier}, \& {Makarov}}]{2004AJ....127.2031K}
{Karachentsev} I.~D., {Karachentseva} V.~E., {Huchtmeier} W.~K., {Makarov}
  D.~I., 2004, \aj, 127, 2031

\bibitem[{{Karachentsev} \& {Nasonova}(2010)}]{2010MNRAS.405.1075K}
{Karachentsev} I.~D., {Nasonova} O.~G., 2010, \mnras, 405, 1075

\bibitem[{{Kitaura}(2013)}]{2013MNRAS.429L..84K}
{Kitaura} F.-S., 2013, \mnras, 429, L84

\bibitem[{{Klypin} {et~al}\mbox{.}(2003){Klypin}, {Hoffman}, {Kravtsov}, \&
  {Gottl{\"o}ber}}]{2003ApJ...596...19K}
{Klypin} A., {Hoffman} Y., {Kravtsov} A.~V., {Gottl{\"o}ber} S., 2003, \apj,
  596, 19

\bibitem[{{Klypin} {et~al}\mbox{.}(2014){Klypin}, {Yepes}, {Gottl\"ober},
  {Prada}, \& {Hess}}]{2014arXiv1411.4001K}
{Klypin} A., {Yepes} G., {Gottl\"ober} S., {Prada} F., {Hess} S., 2014, ArXiv
  e-prints: 1411.4001

\bibitem[{{Klypin} {et~al}\mbox{.}(2011){Klypin}, {Trujillo-Gomez}, \&
  {Primack}}]{Klypin2011}
{Klypin} A.~A., {Trujillo-Gomez} S., {Primack} J., 2011, \apj, 740, 102

\bibitem[{{Knollmann} \& {Knebe}(2009)}]{2009ApJS..182..608K}
{Knollmann} S.~R., {Knebe} A., 2009, \apjs, 182, 608

\bibitem[{{Kolatt} {et~al}\mbox{.}(1996){Kolatt}, {Dekel}, {Ganon}, \&
  {Willick}}]{1996ApJ...458..419K}
{Kolatt} T., {Dekel} A., {Ganon} G., {Willick} J.~A., 1996, \apj, 458, 419

\bibitem[{{Kraan-Korteweg} {et~al}\mbox{.}(1994){Kraan-Korteweg}, {Cayette},
  {Balkowski}, {Fairall}, \& {Henning}}]{1994ASPC...67...99K}
{Kraan-Korteweg} R.~C., {Cayette} V., {Balkowski} C., {Fairall} A.~P.,
  {Henning} P.~A., 1994, 67, 99

\bibitem[{{Kravtsov} {et~al}\mbox{.}(2002){Kravtsov}, {Klypin}, \&
  {Hoffman}}]{2002ApJ...571..563K}
{Kravtsov} A.~V., {Klypin} A., {Hoffman} Y., 2002, \apj, 571, 563

\bibitem[{{Landy} \& {Szalay}(1992)}]{1992ApJ...391..494L}
{Landy} S.~D., {Szalay} A.~S., 1992, \apj, 391, 494

\bibitem[{{Lavaux} \& {Wandelt}(2010)}]{2010MNRAS.403.1392L}
{Lavaux} G., {Wandelt} B.~D., 2010, \mnras, 403, 1392

\bibitem[{{Lee} {et~al}\mbox{.}(1993){Lee}, {Freedman}, \&
  {Madore}}]{1993ApJ...417..553L}
{Lee} M.~G., {Freedman} W.~L., {Madore} B.~F., 1993, \apj, 417, 553

\bibitem[{{Libeskind} {et~al}\mbox{.}(2012){Libeskind}, {Hoffman}, {Knebe},
  {Steinmetz}, {Gottl{\"o}ber}, {Metuki}, \& {Yepes}}]{2012MNRAS.421L.137L}
{Libeskind} N.~I., {Hoffman} Y., {Knebe} A., {Steinmetz} M., {Gottl{\"o}ber}
  S., {Metuki} O., {Yepes} G., 2012, \mnras, 421, L137

\bibitem[{{Ludlow} \& {Porciani}(2011)}]{2011MNRAS.413.1961L}
{Ludlow} A.~D., {Porciani} C., 2011, \mnras, 413, 1961

\bibitem[{{Lynden-Bell} {et~al}\mbox{.}(1988){Lynden-Bell}, {Faber},
  {Burstein}, {Davies}, {Dressler}, {Terlevich}, \&
  {Wegner}}]{1988ApJ...326...19L}
{Lynden-Bell} D., {Faber} S.~M., {Burstein} D., {Davies} R.~L., {Dressler} A.,
  {Terlevich} R.~J., {Wegner} G., 1988, \apj, 326, 19

\bibitem[{{Murray} {et~al}\mbox{.}(2013){Murray}, {Power}, \&
  {Robotham}}]{2013A&C.....3...23M}
{Murray} S.~G., {Power} C., {Robotham} A.~S.~G., 2013, Astronomy and Computing,
  3, 23

\bibitem[{{Nusser} \& {Dekel}(1992)}]{1992ApJ...391..443N}
{Nusser} A., {Dekel} A., 1992, \apj, 391, 443

\bibitem[{{Nusser} {et~al}\mbox{.}(1991){Nusser}, {Dekel}, {Bertschinger}, \&
  {Blumenthal}}]{1991ApJ...379....6N}
{Nusser} A., {Dekel} A., {Bertschinger} E., {Blumenthal} G.~R., 1991, \apj,
  379, 6

\bibitem[{{Planck Collaboration} {et~al}\mbox{.}(2014){Planck Collaboration},
  {Ade}, {Aghanim}, {Armitage-Caplan}, {Arnaud}, {Ashdown}, {Atrio-Barandela},
  {Aumont}, {Baccigalupi}, {Banday}, \& et~al.}]{2014A&A...571A..16P}
{Planck Collaboration} {et~al.}, 2014, \aap, 571, A16

\bibitem[{{Prada} {et~al}\mbox{.}(2012){Prada}, {Klypin}, {Cuesta},
  {Betancort-Rijo}, \& {Primack}}]{Prada2012}
{Prada} F., {Klypin} A.~A., {Cuesta} A.~J., {Betancort-Rijo} J.~E., {Primack}
  J., 2012, \mnras, 423, 3018

\bibitem[{{Sandage}(1994)}]{1994ApJ...430....1S}
{Sandage} A., 1994, \apj, 430, 1

\bibitem[{{Skillman} {et~al}\mbox{.}(2014){Skillman}, {Warren}, {Turk},
  {Wechsler}, {Holz}, \& {Sutter}}]{2014arXiv1407.2600S}
{Skillman} S.~W., {Warren} M.~S., {Turk} M.~J., {Wechsler} R.~H., {Holz} D.~E.,
  {Sutter} P.~M., 2014, ArXiv e-prints: 1407.2600

\bibitem[{{Sorce}(2015)}]{2015MNRAS.450.2644S}
{Sorce} J.~G., 2015, \mnras, 450, 2644

\bibitem[{{Sorce} {et~al}\mbox{.}(2014{\natexlab{a}}){Sorce}, {Courtois},
  {Gottl{\"o}ber}, {Hoffman}, \& {Tully}}]{2014MNRAS.437.3586S}
{Sorce} J.~G., {Courtois} H.~M., {Gottl{\"o}ber} S., {Hoffman} Y., {Tully}
  R.~B., 2014{\natexlab{a}}, \mnras, 437, 3586

\bibitem[{{Sorce} {et~al}\mbox{.}(2013){Sorce}, {Courtois}, {Tully}, {Seibert},
  {Scowcroft}, {Freedman}, {Madore}, {Persson}, {Monson}, \&
  {Rigby}}]{2013ApJ...765...94S}
{Sorce} J.~G. {et~al.}, 2013, \apj, 765, 94

\bibitem[{{Sorce} {et~al}\mbox{.}(2014{\natexlab{b}}){Sorce}, {Tully},
  {Courtois}, {Jarrett}, {Neill}, \& {Shaya}}]{2014MNRAS.444..527S}
{Sorce} J.~G., {Tully} R.~B., {Courtois} H.~M., {Jarrett} T.~H., {Neill} J.~D.,
  {Shaya} E.~J., 2014{\natexlab{b}}, \mnras, 444, 527

\bibitem[{{Springel}(2005)}]{2005MNRAS.364.1105S}
{Springel} V., 2005, \mnras, 364, 1105

\bibitem[{{Stoughton} {et~al}\mbox{.}(2002){Stoughton}, {Lupton}, {Bernardi},
  {Blanton}, {Burles}, {Castander}, {Connolly}, {Eisenstein}, {Frieman},
  {Hennessy}, {Hindsley}, {Ivezi{\'c}}, {Kent}, {Kunszt}, {Lee}, {Meiksin},
  {Munn}, {Newberg}, {Nichol}, {Nicinski}, {Pier}, {Richards}, {Richmond},
  {Schlegel}, {Smith}, {Strauss}, {SubbaRao}, {Szalay}, {Thakar}, {Tucker},
  {Vanden Berk}, {Yanny}, {Adelman}, {Anderson}, {Anderson}, {Annis},
  {Bahcall}, {Bakken}, {Bartelmann}, {Bastian}, {Bauer}, {Berman},
  {B{\"o}hringer}, {Boroski}, {Bracker}, {Briegel}, {Briggs}, {Brinkmann},
  {Brunner}, {Carey}, {Carr}, {Chen}, {Christian}, {Colestock}, {Crocker},
  {Csabai}, {Czarapata}, {Dalcanton}, {Davidsen}, {Davis}, {Dehnen},
  {Dodelson}, {Doi}, {Dombeck}, {Donahue}, {Ellman}, {Elms}, {Evans}, {Eyer},
  {Fan}, {Federwitz}, {Friedman}, {Fukugita}, {Gal}, {Gillespie}, {Glazebrook},
  {Gray}, {Grebel}, {Greenawalt}, {Greene}, {Gunn}, {de Haas}, {Haiman},
  {Haldeman}, {Hall}, {Hamabe}, {Hansen}, {Harris}, {Harris}, {Harvanek},
  {Hawley}, {Hayes}, {Heckman}, {Helmi}, {Henden}, {Hogan}, {Hogg}, {Holmgren},
  {Holtzman}, {Huang}, {Hull}, {Ichikawa}, {Ichikawa}, {Johnston}, {Kauffmann},
  {Kim}, {Kimball}, {Kinney}, {Klaene}, {Kleinman}, {Klypin}, {Knapp},
  {Korienek}, {Krolik}, {Kron}, {Krzesi{\'n}ski}, {Lamb}, {Leger},
  {Limmongkol}, {Lindenmeyer}, {Long}, {Loomis}, {Loveday}, {MacKinnon},
  {Mannery}, {Mantsch}, {Margon}, {McGehee}, {McKay}, {McLean}, {Menou},
  {Merelli}, {Mo}, {Monet}, {Nakamura}, {Narayanan}, {Nash}, {Neilsen},
  {Newman}, {Nitta}, {Odenkirchen}, {Okada}, {Okamura}, {Ostriker}, {Owen},
  {Pauls}, {Peoples}, {Peterson}, {Petravick}, {Pope}, {Pordes}, {Postman},
  {Prosapio}, {Quinn}, {Rechenmacher}, {Rivetta}, {Rix}, {Rockosi}, {Rosner},
  {Ruthmansdorfer}, {Sandford}, {Schneider}, {Scranton}, {Sekiguchi}, {Sergey},
  {Sheth}, {Shimasaku}, {Smee}, {Snedden}, {Stebbins}, {Stubbs}, {Szapudi},
  {Szkody}, {Szokoly}, {Tabachnik}, {Tsvetanov}, {Uomoto}, {Vogeley}, {Voges},
  {Waddell}, {Walterbos}, {Wang}, {Watanabe}, {Weinberg}, {White}, {White},
  {Wilhite}, {Wolfe}, {Yasuda}, {York}, {Zehavi}, \&
  {Zheng}}]{2002AJ....123..485S}
{Stoughton} C. {et~al.}, 2002, \aj, 123, 485

\bibitem[{{Strauss} \& {Willick}(1995)}]{1995PhR...261..271S}
{Strauss} M.~A., {Willick} J.~A., 1995, Physics Reports, 261, 271

\bibitem[{{Teerikorpi}(1990)}]{1990A&A...234....1T}
{Teerikorpi} P., 1990, \aap, 234, 1

\bibitem[{{Teerikorpi}(1993)}]{1993A&A...280..443T}
{Teerikorpi} P., 1993, \aap, 280, 443

\bibitem[{{Teerikorpi}(1995)}]{1995ApL&C..31..263T}
{Teerikorpi} P., 1995, Astrophysical Letters and Communications, 31, 263

\bibitem[{{Teerikorpi}(1997)}]{1997ARA&A..35..101T}
{Teerikorpi} P., 1997, \araa, 35, 101

\bibitem[{{Tinker} {et~al}\mbox{.}(2008){Tinker}, {Kravtsov}, {Klypin},
  {Abazajian}, {Warren}, {Yepes}, {Gottl{\"o}ber}, \&
  {Holz}}]{2008ApJ...688..709T}
{Tinker} J., {Kravtsov} A.~V., {Klypin} A., {Abazajian} K., {Warren} M.,
  {Yepes} G., {Gottl{\"o}ber} S., {Holz} D.~E., 2008, \apj, 688, 709

\bibitem[{{Tonry} {et~al}\mbox{.}(2001){Tonry}, {Dressler}, {Blakeslee},
  {Ajhar}, {Fletcher}, {Luppino}, {Metzger}, \& {Moore}}]{2001ApJ...546..681T}
{Tonry} J.~L., {Dressler} A., {Blakeslee} J.~P., {Ajhar} E.~A., {Fletcher}
  A.~B., {Luppino} G.~A., {Metzger} M.~R., {Moore} C.~B., 2001, \apj, 546, 681

\bibitem[{{Tully}(2010)}]{2010arXiv1010.3787T}
{Tully} R., 2010, ArXiv e-prints: 1010.3787

\bibitem[{{Tully}(2015{\natexlab{a}})}]{2015AJ....149...54T}
{Tully} R.~B., 2015{\natexlab{a}}, \aj, 149, 54

\bibitem[{{Tully}(2015{\natexlab{b}})}]{2015AJ....149..171T}
{Tully} R.~B., 2015{\natexlab{b}}, \aj, 149, 171

\bibitem[{{Tully} {et~al}\mbox{.}(2014){Tully}, {Courtois}, {Hoffman}, \&
  {Pomar{\`e}de}}]{2014Natur.513...71T}
{Tully} R.~B., {Courtois} H., {Hoffman} Y., {Pomar{\`e}de} D., 2014, \nat, 513,
  71

\bibitem[{{Tully} \& {Courtois}(2012)}]{2012ApJ...749...78T}
{Tully} R.~B., {Courtois} H.~M., 2012, \apj, 749, 78

\bibitem[{{Tully} {et~al}\mbox{.}(2013){Tully}, {Courtois}, {Dolphin},
  {Fisher}, {H{\'e}raudeau}, {Jacobs}, {Karachentsev}, {Makarov}, {Makarova},
  {Mitronova}, {Rizzi}, {Shaya}, {Sorce}, \& {Wu}}]{2013AJ....146...86T}
{Tully} R.~B. {et~al.}, 2013, \aj, 146, 86

\bibitem[{{Tully} \& {Fisher}(1977)}]{1977A&A....54..661T}
{Tully} R.~B., {Fisher} J.~R., 1977, \aap, 54, 661

\bibitem[{{Tully} {et~al}\mbox{.}(2008){Tully}, {Shaya}, {Karachentsev},
  {Courtois}, {Kocevski}, {Rizzi}, \& {Peel}}]{2008ApJ...676..184T}
{Tully} R.~B., {Shaya} E.~J., {Karachentsev} I.~D., {Courtois} H.~M.,
  {Kocevski} D.~D., {Rizzi} L., {Peel} A., 2008, \apj, 676, 184

\bibitem[{{Wang} {et~al}\mbox{.}(2014){Wang}, {Mo}, {Yang}, {Jing}, \&
  {Lin}}]{2014ApJ...794...94W}
{Wang} H., {Mo} H.~J., {Yang} X., {Jing} Y.~P., {Lin} W.~P., 2014, \apj, 794,
  94

\bibitem[{{Wang} {et~al}\mbox{.}(2013){Wang}, {Mo}, {Yang}, \& {van den
  Bosch}}]{2013ApJ...772...63W}
{Wang} H., {Mo} H.~J., {Yang} X., {van den Bosch} F.~C., 2013, \apj, 772, 63

\bibitem[{{Watson} {et~al}\mbox{.}(2014){Watson}, {Iliev}, {Diego},
  {Gottl{\"o}ber}, {Knebe}, {Mart{\'{\i}}nez-Gonz{\'a}lez}, \&
  {Yepes}}]{2014MNRAS.437.3776W}
{Watson} W.~A., {Iliev} I.~T., {Diego} J.~M., {Gottl{\"o}ber} S., {Knebe} A.,
  {Mart{\'{\i}}nez-Gonz{\'a}lez} E., {Yepes} G., 2014, \mnras, 437, 3776

\bibitem[{{Willick}(1994)}]{1994ApJS...92....1W}
{Willick} J.~A., 1994, \apjs, 92, 1

\bibitem[{{Willick} {et~al}\mbox{.}(1997){Willick}, {Courteau}, {Faber},
  {Burstein}, {Dekel}, \& {Strauss}}]{1997ApJS..109..333W}
{Willick} J.~A., {Courteau} S., {Faber} S.~M., {Burstein} D., {Dekel} A.,
  {Strauss} M.~A., 1997, \apjs, 109, 333

\bibitem[{{Yepes} {et~al}\mbox{.}(2014){Yepes}, {Gottl{\"o}ber}, \&
  {Hoffman}}]{2014NewAR..58....1Y}
{Yepes} G., {Gottl{\"o}ber} S., {Hoffman} Y., 2014, New Astr. Rev., 58, 1

\bibitem[{{Zaroubi} {et~al}\mbox{.}(1999){Zaroubi}, {Hoffman}, \&
  {Dekel}}]{1999ApJ...520..413Z}
{Zaroubi} S., {Hoffman} Y., {Dekel} A., 1999, \apj, 520, 413

\bibitem[{{Zaroubi} {et~al}\mbox{.}(1995){Zaroubi}, {Hoffman}, {Fisher}, \&
  {Lahav}}]{1995ApJ...449..446Z}
{Zaroubi} S., {Hoffman} Y., {Fisher} K.~B., {Lahav} O., 1995, \apj, 449, 446

\end{thebibliography}

\section*{Appendix}

\begin{figure*}
\includegraphics[width=1 \textwidth]{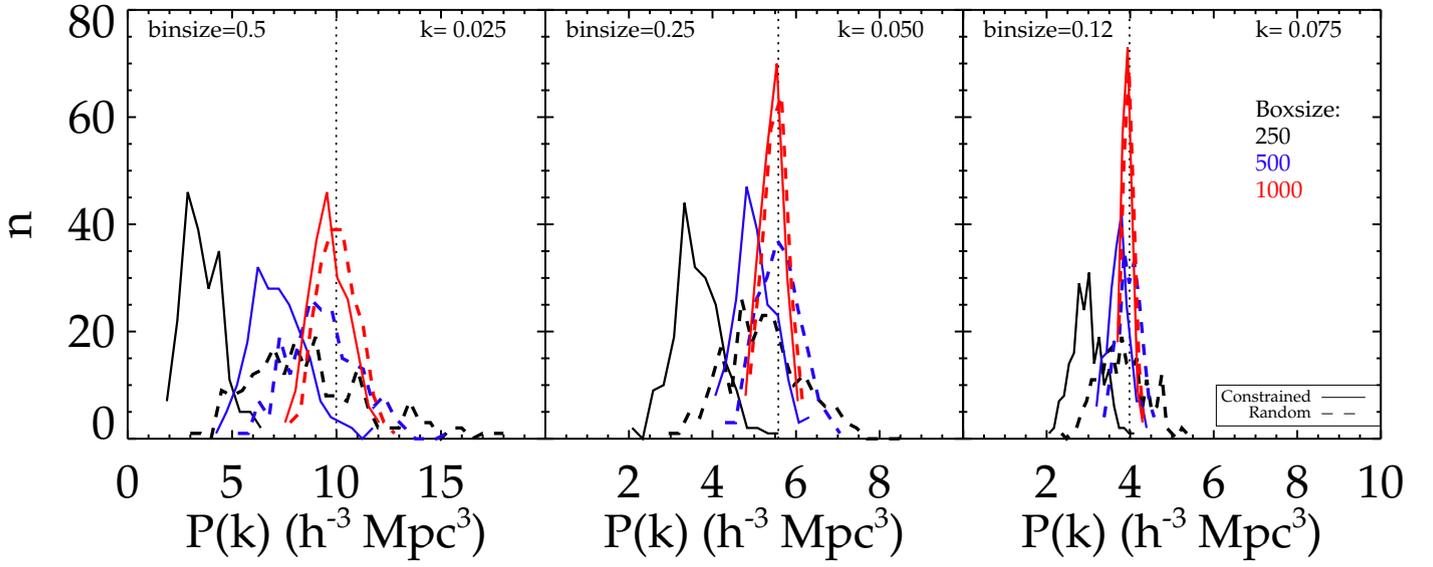}
\vspace{-11cm}
\caption{Distribution of the power spectrum values of constrained (solid line) and random (dashed line) initial conditions at different wavelengths for different boxsizes. From left to right, k=0.025, 0.05 and 0.075 h Mpc$^{-1}$. The black color stand for a boxsize of 250 \hMpc, the blue color for a boxsize of 500 \hMpc\ and the red one for a 1 h$^{-1}$ Gpc boxsize. The dotted black line shows the value of the prior cosmological model power spectrum (Planck in our case) at the corresponding wavelength. }
\label{fig:A}
\end{figure*}

In this appendix, we investigate the effect of the constrained boxsize on the power spectrum at large scales (small wavelengths). To do so, we produce 600 random and 600 constrained initial conditions at low resolution (grid size 128$^3$). One third of these initial conditions is built within a 250 \hMpc\ boxsize, one third within 500 \hMpc\ and the last third is built within a 1 $h^{-1}$ Gpc boxsize. Values of the power spectra of all these initial conditions are then derived at different wavelengths (0.025, 0.05 and 0.075 $h$ Mpc$^{-1}$) and their distributions are plotted as histograms on Figure \ref{fig:A}. From left to right, the wavelength increases (or the scale decreases from $\sim$ 250 to 80 \hMpc) and the dotted line gives the value of the Planck power spectrum at the different wavelengths respectively. We observe that constraining, indeed, tend to decrease the values of the power spectrum (solid lines versus dashed lines) at large scales. As expected, the difference between the power spectrum values of the random and the constrained initial conditions tend to be erased with the increase in boxsize (from black to red), i.e. when smaller and smaller regions of the box are constrained.

\end{document}